\documentclass[aoas]{my_imsart}

\RequirePackage{amsthm,amsmath,amsfonts,amssymb}
\RequirePackage[authoryear]{natbib}
\usepackage{xr-hyper} 
\RequirePackage[colorlinks,citecolor=blue,urlcolor=blue]{hyperref}
\hypersetup{
    colorlinks,
    linkcolor={blue},
    filecolor={blue}, %<----
    urlcolor={blue},
    citecolor={blue}
}
 \usepackage{bm}
 \usepackage{multirow}
 \usepackage{float}
 \usepackage{graphicx}
 \usepackage{mathtools}
 \usepackage{color}

\startlocaldefs
\theoremstyle{remark}
\newtheorem{thm}{Theorem}

\newtheorem{prop}{Proposition}

\newtheorem{setting}{Setting}

\endlocaldefs

\begin{document}

\begin{frontmatter}
\title{An Empirical Bayes Regression for Multi-tissue eQTL Data Analysis}
%\title{A sample article title with some additional note\thanksref{T1}}
\runtitle{Multi-tissue eQTL Data Analysis}
%\thankstext{T1}{A sample of additional note to the title.}

\begin{aug}
%%%%%%%%%%%%%%%%%%%%%%%%%%%%%%%%%%%%%%%%%%%%%%
%%Only one address is permitted per author. %%
%%Only division, organization and e-mail is %%
%%included in the address.                  %%
%%Additional information can be included in %%
%%the Acknowledgments section if necessary. %%
%%%%%%%%%%%%%%%%%%%%%%%%%%%%%%%%%%%%%%%%%%%%%%
%\author[A]{\fnms{First} \snm{Author}\ead[label=e1]{first@somewhere.com}},
\author[A]{\fnms{Fei} \snm{Xue}\ead[label=e3]{feixue@purdue.edu}}
%\author[A]{\fnms{Fei} \snm{Xue}\ead[label=e1,mark]{fei.xue@pennmedicine.upenn.edu}}
\and
\author[B]{\fnms{Hongzhe} \snm{Li}\ead[label=e4]{hongzhe@pennmedicine.upenn.edu}}
%\author[A]{\fnms{Hongzhe} \snm{Li}\ead[label=e2,mark]{hongzhe@pennmedicine.upenn.edu}}
%%%%%%%%%%%%%%%%%%%%%%%%%%%%%%%%%%%%%%%%%%%%%%
%% Addresses                                %%
%%%%%%%%%%%%%%%%%%%%%%%%%%%%%%%%%%%%%%%%%%%%%%
%\address[A]{Department,
%University or Company Name,
%\printead{e1}}

\address[A]{%Department of Statistics, 
Purdue University,  
\printead{e3}
}

\address[B]{%Department of Biostatistics, Epidemiology and Informatics, 
University of Pennsylvania, 
\printead{e4}}

%\author[A]{\fnms{Fei} \snm{Xue}\ead[label=e1,mark]{fei.xue@pennmedicine.upenn.edu}}
%%\and
%\author[A]{\fnms{Hongzhe} \snm{Li}\ead[label=e2,mark]{hongzhe@pennmedicine.upenn.edu}}
%%\author[B]{\fnms{???} \snm{???}\ead[label=e3,mark]{???@???}}
%%%%%%%%%%%%%%%%%%%%%%%%%%%%%%%%%%%%%%%%%%%%%%%
%%% Addresses                                %%
%%%%%%%%%%%%%%%%%%%%%%%%%%%%%%%%%%%%%%%%%%%%%%%
%\address[A]{Department of Biostatistics, Epidemiology and Informatics, University of Pennsylvania, \printead{e1, e2}}
%
%%\address[B]{???, \printead{e2,e3}}

\end{aug}

\begin{abstract}
The Genotype-Tissue Expression (GTEx)  project  collects samples from multiple human tissues to study the relationship between genetic variation or single nucleotide polymorphisms (SNPs) and gene expression in each tissue. However, most existing eQTL analyses only focus on single tissue information. In this paper, we develop a multi-tissue eQTL analysis that improves the  single tissue cis-SNP gene expression association analysis by borrowing information across tissues. Specifically, we propose an empirical Bayes regression model for SNP-expression association analysis using data across multiple tissues. To allow the effects of SNPs to vary greatly among tissues, we use a mixture distribution as the prior, which is a mixture of a multivariate Gaussian distribution and a Dirac mass at zero. The model allows us to assess the cis-SNP gene expression association in each tissue  by calculating the Bayes factors. We show that the proposed estimator of the cis-SNP effects on gene expression achieves the minimum Bayes risk among all estimators. Analyses of the GTEx data show that our proposed method is superior to traditional simple regression methods in terms of predicting  accuracy for gene expression levels using cis-SNPs in testing data sets. Moreover, we find that although genetic effects on expression are extensively shared among tissues, effect sizes still vary greatly across tissues.
\end{abstract}

\begin{keyword}
\kwd{Bayes risk}
\kwd{Data integration}
\kwd{EM algorithm}
\kwd{GTEx}
\kwd{Missing data}
\kwd{Mixture model}
\end{keyword}

\end{frontmatter}
%%%%%%%%%%%%%%%%%%%%%%%%%%%%%%%%%%%%%%%%%%%%%%
%% Please use \tableofcontents for articles %%
%% with 50 pages and more                   %%
%%%%%%%%%%%%%%%%%%%%%%%%%%%%%%%%%%%%%%%%%%%%%%
%\tableofcontents

%%%%%%%%%%%%%%%%%%%%%%%%%%%%%%%%%%%%%%%%%%%%%%
%%%% Main text entry area:

\section{Introduction}
Expression quantitative trait loci (eQTL) analysis aims to identify single nucleotide polymorphisms (SNPs) that are associated with the expression of a gene in a given tissue.  Many large data sets have been generated for such genetics of gene expression studies for various tissues, which have provided important insights into gene regulations. Among these studies, the Genotype-Tissue Expression (GTEx) project aims to characterize variation in gene expression levels across individuals and diverse tissues of the human body, many of which are not easily accessible \citep{gtex}.  The project found that  local genetic variation affects gene expression levels for the majority of genes, and  identified  inter-chromosomal genetic effects for a small number of genes and loci.  
Such eQTL analyses provide important insights into genetic regulation of gene expressions.  The GTEx data sets have also been applied to impute gene expression levels based on genetic variants data and the imputed gene expressions were subsequently used in transcriptome-wide  association analysis such as in TWAS analysis \citep{gamazon2015gene, gusev2016integrative, hu2019statistical}.

However, small sample sizes of many eQTL studies, including the GTEx data, often have limited power to detect eQTLs  and large variance in estimated gene expression levels based on genotype data. To date, most eQTL studies have considered the association  between genetic variation and expression in a single tissue \citep{brem2005genetic, stranger2007population, stegle2012using}. Multi-tissue eQTL analysis has the potential to improve the findings of single tissue analyses by borrowing strength across tissues, and the potential to elucidate the genetic basis of the difference in expressions  between tissues. \cite{GenLi} proposed a multivariate hierarchical Bayesian model (MT-eQTL) for multi-tissue eQTL analysis.  The MT-eQTL directly models the vector of correlations between expression and genotype across tissues.  \cite{Stephens} developed another mixture-model based approach and showed that  their analysis identifies more eQTLs than existing approaches, consistent with improved power. 
 \cite{sul2013effectively} adopted a linear mixed model to detect eQTLs across multiple tissues.
 In addition, \cite{duong2017applying} proposed a meta-analysis model with random effects to combine eQTL studies from different tissues. However, these methods only focus on estimation of association between gene expression and one single SNP.
 
 Instead of considering a SNP-gene pair in the single-locus models,
% typical eQTL data analysis, 
% several methods 
 multi-locus models were developed recently for the joint effects of  multiple SNPs on gene expressions, which incorporate the correlations between the multiple SNPs.  %\cite{sul2015accurate} propose an efficient approach for multiple tests of multiple SNPs. 
 For example, \cite{zeng2017cis} use a linear mixed model and a likelihood ratio test to examine whether multiple SNPs are jointly associated with a gene. \cite{gosik2017iform} adopts a high-dimensional regression variable selection model to jointly analyze multiple SNPs at one time. In addition, \cite{bhadra2013joint} use a high-dimensional Bayesian variable selection model for association between multiple SNPs and multiple gene expression responses.
 Nevertheless, these methods do not  consider the shared information across multiple tissues.

In this paper, we develop an empirical Bayes regression model for SNP-expression association analysis using data across multiple tissues, allowing for considering the joint effects of multiple genetic variants on gene expressions.  Specifically, our model serves two purposes. One  
is to estimate association  the between gene expression and the corresponding cis-SNPs for each gene and tissue, where cis-SNPs of a given gene are SNPs located either within the gene or in the $1$ Mb upstream and downstream regions of the gene.
The other is to test whether these cis-SNPs are associated with the gene expression, that is, whether the gene is an eGene whose expression level is related to at least one cis-SNP \citep{duong2016using}. 

To achieve these goals, for each gene, we construct tissue-specific linear regression models with the gene expression level of the gene as the response and its corresponding cis-SNPs as predictors. To borrow information across different tissues, we propose an empirical Bayes estimator for the regression coefficients based on a mixture prior distribution. Specifically, we adopts the posterior mean of the coefficients for estimation. In addition, we estimate the prior parameters in the the posterior mean through maximizing the marginal likelihood of the gene expression values in all tissues based on the expectation–maximization (EM) algorithm \citep{dempster1977maximum}. Moreover,  we extract evidence of  whether the cis-SNPs are relevant to the gene expression from the data by
calculating posterior probabilities and Bayes factor of hypotheses.

The main contributions of the proposed method are as follows.
First, we extract shared information across tissues via the common prior distribution of the regression coefficients in different tissue-specific models. We propose to combine the information in each single tissue and the shared information in the prior distribution using the empirical Bayes estimator.
Compared with the traditional ordinary least squared (OLS) estimator of the coefficients in each tissue-specific model, the proposed empirical Bayes estimator improves the estimation of the regression coefficients in terms of the Bayes risk in Section \ref{sec: theory} and the mean squared error.
Second, we incorporate situations where all the  cis-SNPs are irrelevant to the gene expression in some tissues, since our assigned prior distribution of coefficients is a mixture with two components. One of the two components is exactly a zero vector, while the other one has non-zero mean representing shared information across tissues that have non-zero effects. In this way, we can test whether a gene is an eGene in a specific tissue based on the posterior probabilities of the assignments for the two components for a  given tissue.
Through our analysis of the GTEx data in Section \ref{sec: real}, we show that while genetic effects on expression are extensively shared among tissues, effect sizes can still vary greatly among the  tissues.

%The rest of the paper  are organized as follows. We propose the empirical Bayes model in Section \ref{sec: EB}. In Section \ref{sec: algorithm}, the EM algorithm are illustrated. We establish theoretical properties of the proposed estimator in Section \ref{sec: theory}. Sections
% \ref{sec: sim} and \ref{sec: real} demonstrate that the proposed method performs better than the traditional OLS through both simulations and the  GTEx data application.
%Specifically, we develop an empirical Bayes method to borrow information across tissues.

%Our goal is to estimate relationship between gene expression values and genotype data across multiple tissues based on samples in the Genotype-Tissue Expression (GTEx) data. 

\section{An empirical Bayes regression model for SNP-expression association across multiple tissues} \label{sec: EB}
\subsection{Empirical Bayes regression} \label{sec: model}
%Let $n$ be the number of total individuals and $m$ be the number of tissues in the data. Let $\bm{Y}$  be a $n\times m$ matrix consisting of expression values of a gene in all tissues, $\bm{Y}^{(t)}$ be the $t$-th column in $\bm{Y}$, and $\bm{X}$ be a $n\times p$ constant matrix consisting of cis-SNPs for the gene. 
%We assume the following tissue-specific linear regression model to link the SNP genotypes with gene expression, 

In this section, we link the SNP genotypes with gene expression in each tissue by a tissue-specific linear regression model. Specifically, we let $\bm{Y}$ denote a $n\times m$ matrix consisting of expression values of a gene  in  $m$ tissues of $n$ samples and $\bm{X}$ denote a $n\times p$ constant matrix consisting of cis-SNPs for the gene, where $n$ is the number of total individuals, $m$ is the number of tissues in the data, and $p$ is the number of cis-SNPs. For the $t$-th tissue, the tissue-specific linear regression model is
\begin{equation}\label{model}
\bm{Y}^{(t)}=\bm{X}\bm{\beta}^{(t)}+\bm{\varepsilon}^{(t)},
\end{equation}
where $\bm{Y}^{(t)}$ denotes the $t$-th column in $\bm{Y}$,
$\bm{\beta}^{(t)}$ is a $p$-dimensional coefficient vector for the $t$-th tissue, and $\bm{\varepsilon}^{(t)}\sim N_n(\bm{0}, \sigma^2\bm{I}_n)$ is the error term with parameter $\sigma>0$ and independent of $\bm{X}\bm{\beta}^{(t)}$.
We also assume that $\bm{\varepsilon}^{(t)}$ for $t=1, \dots, m$ are independent.
 The 
ordinary least squares (OLS) estimator of $\bm{\beta}^{(t)}$ based on information in a single tissue is
\begin{eqnarray}\label{OLS}
\hat{\bm{\beta}}^{(t)}=(\bm{X}^T  \bm{X})^{-1}\bm{X}^T  \bm{Y}^{(t)}=\bm{\beta}^{(t)}+(\bm{X}^T  \bm{X})^{-1}\bm{X}^T  \bm{\varepsilon}^{(t)},
\end{eqnarray}
Then $\hat{\bm{\beta}}^{(t)}\mid\bm{\beta}^{(t)} \sim N_{p} (\bm{\beta}^{(t)}, \sigma^2(\bm{X}^T  \bm{X})^{-1})$.

To borrow information across tissues, we assume that coefficient vectors $\bm{\beta}^{(t)}$ over all the tissues are random and have a common prior distribution. This common prior contains shared effects of cis-SNPs across tissues. {However, the effects of cis-SNPs in different tissues could vary greatly. Especially, the cis-SNPs could be ``inactive'' and have no effects on gene expression in some tissues, which can not contribute to the shared effects.
To accommodate this possibility,} we define a random indicator $I^{(t)}$, which follows a Bernoulli distribution with probability $\tau_1 \in(0,1)$, to reflect the status of $\bm{\beta}^{(t)}$.
 We assign a mixture prior distribution with two mixture components 
% multivariate normal prior 
 for 
 $\bm{\beta}^{(t)}$, %under $I^{(t)}=1$, 
 that is,
 \begin{eqnarray}\label{models}
 &&\bm{\beta}^{(t)} \mid I^{(t)}=1 \sim N_{p} (\bm{\beta}, \eta (\bm{X}^T\bm{X})^{-1}),\\
 &&\bm{\beta}^{(t)} \mid I^{(t)}=0 \equiv \bm{0}, \label{model2}
 \end{eqnarray}
 independently for $t=1, \dots, m$, where $\eta>0$ is a parameter.  
 Here $I^{(t)}$ is a latent configuration variable determining the assignment of $\bm{\beta}^{(t)}$ to a mixture component in the mixture prior distribution. When $I^{(t)}=0$, the cis-SNPs are ``inactive'' and have no effects on the gene expression $\bm{Y}^{(t)}$. In contrast, when $I^{(t)}=1$, the cis-SNPs are ``active'' and the effects $\bm{\beta}^{(t)}$ follows a multivariate normal distribution with mean $\bm{\beta}$.
 
The proposed method models the relationship between the gene expression and all the cis-SNPs jointly, while the related existing methods proposed by \cite{GenLi} and \cite{Stephens} model the relationship between gene expression and one cis-SNP individually. In addition, the proposed method borrows cross-tissue information via the prior distribution $N_{p} (\bm{\beta}, \eta (\bm{X}^T\bm{X})^{-1})$, while the existing MT-eQTL method in \cite{GenLi} uses a covariance matrix in a prior distribution to capture correlations between any two tissues. Moreover, the proposed method and the MT-eQTL method are empirical Bayes approaches, while \cite{Stephens} adopt a full Bayes framework.
 
We provide the posterior probabilities of $I^{(t)}$ and the posterior mean of $\bm{\beta}^{(t)}$ in the following proposition. Let $\psi(\bm{z}; \bm{\mu}_0, \bm{\Sigma}_0)$ denote the density function of the multivariate normal distribution $N (\bm{\mu}_0, \bm{\Sigma}_0)$.

  %\textcolor{blue}{
 % 	Let $\psi(\bm{z}; \bm{\mu}_0, \bm{\Sigma}_0)$ represent the density function of $N_{p} (\bm{\mu}_0, \bm{\Sigma}_0)$. Let $\bm{H}=\bm{X} (\bm{X}^T\bm{X})^{-1}\bm{X}^T$.
  %}

%{\color{blue}
	\begin{prop}\label{pm}
		The posterior means of $\bm{\beta}^{(t)}$ given $I^{(t)}$ are
		$$E(\bm{\beta}^{(t)} \mid  \bm{Y}^{(t)},  I^{(t)}=1)=\left(\frac{1}{\eta} +\frac{1}{\sigma^2}\right)^{-1} \left(\frac{\bm{\beta}}{\eta} + \frac{\hat{\bm{\beta}}^{(t)}}{\sigma^2}\right),$$
		and $E(\bm{\beta}^{(t)} \mid  \bm{Y}^{(t)}, I^{(t)}=0)=0$.
		The posterior probabilities of $I^{(t)}$ are 
		$$P(I^{(t)} =1  \mid \bm{Y} )= h_1(\bm{Y}^{(t)}; \tau_1, \bm{\beta}, \eta, \sigma^2),$$
		and $P(I^{(t)} =0  \mid \bm{Y}) = 1-h_1(\bm{Y}^{(t)}; \tau_1, \bm{\beta}, \eta, \sigma^2),$
		where
		\begin{equation*}
		h_1(\bm{Y}^{(t)}; \tau_1, \bm{\beta}, \eta, \sigma^2)=\frac{\tau_1 \psi(\bm{Y}^{(t)}; \bm{X} \bm{\beta}, \sigma^2 \bm{I}_{n}+ \eta\bm{H})}{\tau_1 \psi(\bm{Y}^{(t)}; \bm{X} \bm{\beta}, \sigma^2 \bm{I}_{n}+ \eta\bm{H})+ \tau_0 \psi(\bm{Y}^{(t)}; \bm{0}, \sigma^2 \bm{I}_{n})},
		\end{equation*}
		 with $\bm{H}=\bm{X} (\bm{X}^T\bm{X})^{-1}\bm{X}^T$ and $\tau_0=1-\tau_1$.
		Thus, the posterior mean of $\bm{\beta}^{(t)}$ is
		\begin{eqnarray}\label{postmean}
			E(\bm{\beta}^{(t)} \mid  \bm{Y})
			&=&E(\bm{\beta}^{(t)} \mid  \bm{Y}^{(t)})
			=h_1(\bm{Y}^{(t)}; \tau_1, \bm{\beta}, \eta, \sigma^2) \left(\frac{1}{\eta} +\frac{1}{\sigma^2}\right)^{-1} \left(\frac{\bm{\beta}}{\eta} + \frac{\hat{\bm{\beta}}^{(t)}}{\sigma^2}\right).
		\end{eqnarray}
	\end{prop}
%}

According to the Proposition \ref{pm}, the posterior mean of $\bm{\beta}^{(t)}$ is a weighted average of the OLS estimator in Equation (\ref{OLS}) and the mean of the prior distribution in Equation (\ref{models}), which combines the information in the $t$-th tissue and the shared information across tissues in the prior.
The first equality in (\ref{postmean}) follows from the fact that $\bm{\beta}^{(t)}$ and columns in $\bm{Y}$ other than $\bm{Y}^{(t)}$ are conditionally independent given $\bm{Y}^{(t)}$.
The weights in (\ref{postmean}) are related to the variance of the error term and the variance of the prior distribution.

%{\color{red}
\subsection{Bayes factor}\label{sec: BF}
%}
For a given tissue $t$, to determine whether  the cis-SNPs is relevant to the  gene expression, we can calculate the posterior probability $p(I^{(t)} = 0|\bm{Y})$ or the Bayesian factor. 
Specifically, we access the plausibility of the cis-SNP and gene expression association  via
the Bayes factor (BF) 
\begin{eqnarray}\label{BF}
K_t(\bm{Y}; \bm{\beta}, \eta, \sigma^2)&=&\frac{p(\bm{Y} \mid H_0^{(t)})}{p(\bm{Y}  \mid H_1^{(t)})} 
= \frac{p(\bm{Y}^{(t)} \mid I^{(t)}=0)}{p(\bm{Y}^{(t)} \mid I^{(t)} =1)}
%&=&\frac{\int p(\bm{\beta}^{(t)}\mid I^{(t)}=0) p(\bm{Y}^{(t)} \mid I^{(t)}=0, \bm{\beta}^{(t)}) d\bm{\beta}^{(t)}}{\int p(\bm{\beta}^{(t)}\mid I^{(t)}=1) p(\bm{Y}^{(t)} \mid I^{(t)}=1, \bm{\beta}^{(t)}) d\bm{\beta}^{(t)}}\notag\\
%&=&\frac{ p(\bm{W}^{(t)})  \psi(\bm{Y}^{(t)}; \bm{0}, \sigma^2 \bm{I})}{p(\bm{W}^{(t)}) \psi(\bm{Y}^{(t)}; \bm{X} \bm{\beta}, \sigma^2 \bm{I} + \eta\bm{H})}\notag\\
%&=&\frac{f(\hat{\bm{\beta}}^{(t)}; \bm{0},  \sigma^2(\bm{X}^T\bm{X})^{-1})}{\int f(\bm{\beta}^{(t)}; \bm{\beta},  \eta (\bm{X}^T\bm{X})^{-1}) f(\hat{\bm{\beta}}^{(t)}; \bm{\beta}^{(t)},  \sigma^2(\bm{X}^T\bm{X})^{-1}) d\bm{\beta}^{(t)}}\notag\\
=\frac{\psi(\bm{Y}^{(t)}; \bm{0}, \sigma^2 \bm{I})}{\psi(\bm{Y}^{(t)}; \bm{X} \bm{\beta}, \sigma^2 \bm{I}+\eta\bm{H})},
\end{eqnarray}
based on the prior distribution in Equations (\ref{models}) and (\ref{model2}).
The second equality in (\ref{BF}) is due to that columns in $\bm{Y}$ other than $\bm{Y}^{(t)}$ do not depend on $I^{(t)}$, which implies that $K_t(\bm{Y}; \bm{\beta}, \eta, \sigma^2)=K_t(\bm{Y}^{(t)}; \bm{\beta}, \eta, \sigma^2)$.
In addition, the posterior odds ratio is
\begin{equation}\label{postodds}
O_t(\bm{Y}; \tau_1, \bm{\beta}, \eta, \sigma^2)=\frac{P(I^{(t)}=0 \mid \bm{Y}^{(t)})}{P(I^{(t)} =1  \mid \bm{Y}^{(t)})} = K_t(\bm{Y}; \bm{\beta}, \eta, \sigma^2)\cdot \frac{\tau_0}{\tau_1}.
\end{equation}
%\begin{eqnarray}\label{postodds}
%O_t(\bm{Y}; \tau_1, \bm{\beta}, \eta, \sigma^2)&=&\frac{P(I^{(t)}=0 \mid \bm{Y}^{(t)})}{P(I^{(t)} =1  \mid \bm{Y}^{(t)})}\notag\\
%&=&K_t(\bm{Y}; \bm{\beta}, \eta, \sigma^2)\cdot \frac{P(I^{(t)} =0)}{P(I^{(t)} =1)} = K_t(\bm{Y}; \bm{\beta}, \eta, \sigma^2)\cdot \frac{1-\tau_1}{\tau_1}.
%\end{eqnarray}
From equation (\ref{postodds}), we see that the BF can indicate whether the observed data provides evidence for or against the cis-SNP gene expression association.  If BF $> 1$ then the posterior odds are greater than the prior odds $\tau_0/\tau_1$, indicating that the observed data provides evidence for the no cis-SNP gene expression association.  If BF $< 1$ then the data provides evidence for  cis-SNP gene expression  association.  
%posterior odds are less than the prior odds.
We estimate the unknown parameters in Equations (\ref{postmean}) and (\ref{BF}) via an expectation-maximization (EM) algorithm \citep{dempster1977maximum} in the following section.

\section{Parameter estimation and EM algorithm}\label{sec: algorithm}

In this section, we provide a detailed iterative algorithm to estimate the parameters in the mixture prior distribution through maximizing the likelihood of the data. 
Specifically, we exploit an EM algorithm \citep{dempster1977maximum} 
to find the maximum likelihood estimate (MLE) of $\bm{\theta}$, where $\bm{\theta}$ consists of all the parameters in the model, that is, $\bm{\theta}=(\tau_1, \tau_0, \bm{\beta}, \eta, \sigma^2)$.
Each iteration consists of an expectation step and a maximization step. 
Suppose that we have both $\bm{Y}^{(t)}$ and $\bm{I}^{(t)}$ for each $t=1, \dots, m$. We refer to $\{\bm{Y}, I^{(1)}, \dots, I^{(m)}\}$ as the complete data. %and the incomplete data, respectively.
The complete-data likelihood is 
\begin{eqnarray*}
p(\bm{Y}, I^{(1)}, \dots, I^{(m)}; \tau_1, \tau_0, \bm{\beta}, \eta, \sigma^2)
=\prod_{t=1}^{m}\prod_{s=0}^{1}\left\{\tau_s g_s(\bm{Y}^{(t)}; \bm{\beta}, \eta, \sigma^2)\right\}^{\mathbb{I}(I^{(t)}=s)},
\end{eqnarray*}
where $\mathbb{I}(\cdot)$ is an indicator function, %$\tau_0=1-\tau_1$, 
$g_0(\bm{Y}^{(t)}; \bm{\beta}, \eta, \sigma^2)=g_0(\bm{Y}^{(t)}; \sigma^2)= \psi(\bm{Y}^{(t)}; \bm{0}, \sigma^2 \bm{I}_{n})$ 
and 
$g_1(\bm{Y}^{(t)}; \bm{\beta}, \eta, \sigma^2)= \psi(\bm{Y}^{(t)}; \bm{X} \bm{\beta}, \sigma^2 \bm{I}_{n} + \eta\bm{H})$ denote the likelihoods of $\bm{Y}^{(t)}$ when $I^{(t)}=0$ and $I^{(t)}=1$, respectively.
In the expectation step, since we typically do not observe $\{I^{(1)}, \dots, I^{(m)}\}$ in practice,
given the current estimate $\bm{\theta}_{(k)}$ at the $k$-th iteration,
 we first calculate the posterior distribution of $I^{(t)}$  
\begin{eqnarray*}
T^{(t)}_{s,(k)} = P(I^{(t)}=s \mid \bm{Y}, \bm{\theta}_{(k)})
=\frac{\tau_{s,(k)} g_s(\bm{Y}^{(t)}; \bm{\beta}_{(k)}, \eta_{(k)}, \sigma_{(k)})}{\tau_{1,(k)} g_1(\bm{Y}^{(t)}; \bm{\beta}_{(k)}, \eta_{(k)}, \sigma_{(k)})+\tau_{0,(k)} g_0(\bm{Y}^{(t)}; \sigma_{(k)})}
\end{eqnarray*}
%represent posterior probabilities of $I^{(t)}$ given $\bm{\theta}_{(k)}$,
for $s=0,1$, where $\tau_{s,(k)}$, $\bm{\beta}_{(k)}$, $\eta_{(k)}$, and $\sigma_{(k)}$ denote estimates of $\tau$, $\bm{\beta}$, $\eta$, and $\sigma$ at the $k$-th iteration.
%Since we do not observe the $\{I^{(0)}, \dots, I^{(m)}\}$, 
Moreover,  
%(E step): 
%for each $s=0,1$,
%\begin{eqnarray}
%T^{(t)}_{s,(k)}&\defeq & P(I^{(t)}=s \mid \bm{Y}^{(t)}, \bm{\theta}_{(k)})\\
%&=&\frac{\tau_{s,(k)} g_s(\bm{Y}^{(t)}; \bm{\beta}_{(k)}, \eta_{(k)}, \sigma_{(k)})}{\tau_{1,(k)} g_1(\bm{Y}^{(t)}; \bm{\beta}_{(k)}, \eta_{(k)}, \sigma_{(k)})+\tau_{0,(k)} g_0(\bm{Y}^{(t)}; \sigma_{(k)})}.
%\end{eqnarray}
%note that
we calculate the expectation of the complete-data log-likelihood under the posterior distribution of the latent variables $\{I^{(1)}, \dots, I^{(m)}\}$:
%based on $\bm{\theta}_{(k)}$
\begin{eqnarray*}
Q(\bm{\theta}\mid\bm{\theta}_{(k)})
&=&E_{I^{(1)}, \dots, I^{(m)} \mid \bm{Y}, \bm{\theta}_{(k)}}\left[\log p(\bm{Y}, I^{(1)}, \dots, I^{(m)};  \bm{\theta}
%\tau_1, \tau_0, \bm{\beta}, \eta, \sigma^2
)\right]\\
%&=&\sum_{t=1}^{m} \sum_{s=0}^{1} P(I^{(t)}=s \mid \bm{Y}^{(t)}, \bm{\theta}_{(k)}) \left\{ \log \tau_s + \log g_s(\bm{Y}^{(t)}; \bm{\beta}, \eta, \sigma^2)\right\}\\
&=&\sum_{t=1}^{m} \sum_{s=0}^{1} T^{(t)}_{s,(k)} \left\{ \log \tau_s + \log g_s(\bm{Y}^{(t)}; \bm{\beta}, \eta, \sigma^2)\right\}.
\end{eqnarray*}

In the maximization step, %(M step):
we maximize this expectation to determine the next estimate for all the parameters.
%We compute the next estimate of $\bm{\theta}$, which is 
The
 maximizer of $Q(\bm{\theta}\mid\bm{\theta}_{(k)})$ consists of
 \begin{equation*}
\tau_{s,(k+1)}=\frac{\sum_{t=1}^{m} T^{(t)}_{s,(k)}}{\sum_{t=1}^{m} \left\{ T^{(t)}_{0,(k)} +T^{(t)}_{1,(k)} \right\} } \quad \text{ for } s=1,0, \quad
%\tau_{1,(k+1)}=\frac{\sum_{t=1}^{m} T^{(t)}_{1,(k)}}{\sum_{t=1}^{m} \left\{ T^{(t)}_{0,(k)} +T^{(t)}_{1,(k)} \right\} }, 
\bm{\beta}_{(k+1)}=\frac{\sum_{t=1}^{m} T^{(t)}_{1,(k)}\hat{\bm{\beta}}^{(t)}}{\sum_{t=1}^{m} T^{(t)}_{1,(k)}},
\end{equation*}
\begin{equation*}
 \sigma_{(k+1)}^2=\frac{\sum_{t=1}^{m}\left(\bm{Y}^{(t)}\right)^T \left(\bm{Y}^{(t)}\right) - \sum_{t=1}^{m}T^{(t)}_{1,(k)} \left(\bm{Y}^{(t)}\right)^T \bm{H}\left(\bm{Y}^{(t)}\right) }{mn - p \sum_{t=1}^{m}T^{(t)}_{1,(k)}},
\end{equation*}
and
\begin{equation*}
\eta_{(k+1)}=\frac{\sum_{t=1}^{m}T^{(t)}_{1,(k)} \left(\bm{Y}^{(t)}-\bm{X} \bm{\beta}_{(k+1)}\right)^T \bm{H}\left(\bm{Y}^{(t)}-\bm{X} \bm{\beta}_{(k+1)}\right) }{p\sum_{t=1}^{m}T^{(t)}_{1,(k)}}-\sigma_{(k+1)}^2.
\end{equation*}
In this way, we derive closed-form expression updates for each iteration, which are straightforward to compute. In addition, this EM algorithm converges since each iteration does increase the likelihood of observed data.

\section{Statistical properties} \label{sec: theory}

%With the covariance matrix assumption $\bm{\Sigma}=\eta (\bm{X}^T\bm{X})^{-1}$, 

%where $\hat{\bm{B}}^{(m)}=(\hat{\bm{\beta}}^{(1)}, \dots, \hat{\bm{\beta}}^{(m)})$, and
%\begin{equation*}
%h_1(\hat{\bm{\beta}}^{(t)}; \widetilde{\tau}_1, \widetilde{\bm{\beta}}, \widetilde{\eta}, \widetilde{\sigma}) = \frac{\widetilde{\tau}_1 f(\hat{\bm{\beta}}^{(t)}; \widetilde{\bm{\beta}}, (\widetilde{\eta}+\widetilde{\sigma}^2)(\bm{X}^T\bm{X})^{-1})}{\widetilde{\tau}_1 f(\hat{\bm{\beta}}^{(t)}; \widetilde{\bm{\beta}}, (\widetilde{\eta}+\widetilde{\sigma}^2)(\bm{X}^T\bm{X})^{-1})+(1-\widetilde{\tau}_1) f(\hat{\bm{\beta}}^{(t)}; \bm{0},  \widetilde{\sigma}^2(\bm{X}^T\bm{X})^{-1})}.
%\end{equation*}

In this section, we provide the asymptotic results of the proposed estimator %and an oracle estimator 
in terms of a Bayes risk function. Specifically, we define the Bayes risk function of an estimator $\bm{\delta}_m(\bm{Y}) \in \mathcal{E}_m$ for $\bm{\beta}^{(t)}$ as
\begin{equation*}
R_m(\bm{\delta}_m)=\int l\left(\bm{\beta}^{(t)}, \bm{\delta}_m(\bm{Y})\right) \prod_{i=1}^{m} \left\{ p(\bm{Y}^{(i)} \mid \bm{\beta}^{(i)}) p(\bm{\beta}^{(i)}) d\bm{Y}^{(i)} d\bm{\beta}^{(i)} \right\},
\end{equation*}
where $\mathcal{E}_m$ is the set consisting of all available estimators of $\bm{\beta}^{(t)}$, and
\begin{equation*}
l\left(\bm{\beta}^{(t)}, \widetilde{\bm{\beta}}^{(t)}(\bm{Y})\right) = \left\{\widetilde{\bm{\beta}}^{(t)} (\bm{Y}) - \bm{\beta}^{(t)}\right\}^T\Delta \left\{\widetilde{\bm{\beta}}^{(t)} (\bm{Y}) - \bm{\beta}^{(t)}\right\},
\end{equation*}
is a squared error loss function with a positive definite matrix $\Delta$. 
Let $\widetilde{\tau}_0$, $\widetilde{\tau}_1$, $\widetilde{\eta}$, $\widetilde{\sigma}$, and $\widetilde{\bm{\beta}}$ be %the solution of the EM algorithm, that is, 
the
%maximizer of equation (\ref{marginal_y}), that is, $\widetilde{\tau}_1$, $\widetilde{\eta}$, $\widetilde{\sigma}$, and $\widetilde{\bm{\beta}}$ are 
MLEs of $\tau_0$, $\tau_1$, $\eta$, $\sigma$, and $\bm{\beta}$, respectively. Then, by Proposition \ref{pm}, the proposed empirical Bayes estimator of $\bm{\beta}^{(t)}$ for the $t$-th tissue is 
\begin{eqnarray*}
\widetilde{\bm{\beta}}^{(t)} (\bm{Y}) %&=& h_1(\hat{\bm{\beta}}^{(t)}; \widetilde{\tau}_1, \widetilde{\bm{\beta}}, \widetilde{\eta}, \widetilde{\sigma})\left(\frac{1}{\widetilde{\eta}}\bm{X}^T\bm{X}+\frac{1}{\widetilde{\sigma}^2}\bm{X}^T\bm{X}\right)^{-1}\left(\frac{1}{\widetilde{\eta}}\bm{X}^T\bm{X}\widetilde{\bm{\beta}} + \frac{1}{\widetilde{\sigma}^2} \bm{X}^T\bm{X}\hat{\bm{\beta}}^{(t)}\right)\\
&=&\frac{\widetilde{\eta} \widetilde{\sigma}^2 \cdot h_1(\bm{Y}^{(t)}; \widetilde{\tau}_1, \widetilde{\bm{\beta}}, \widetilde{\eta}, \widetilde{\sigma})}{\widetilde{\eta}+\widetilde{\sigma}^2}\left(\frac{\widetilde{\bm{\beta}}}{\widetilde{\eta}} + \frac{\hat{\bm{\beta}}^{(t)}}{\widetilde{\sigma}^2}\right).
\end{eqnarray*}
If $\tau_1$, $\eta$, $\sigma$, and $\bm{\beta}$ are known, then we can use
\begin{eqnarray}\label{oracle}
\bar{\bm{\beta}}^{(t)} (\bm{Y})=E(\bm{\beta}^{(t)} \mid \bm{Y}^{(t)}) 
=\frac{\eta \sigma^2 \cdot h_1(\bm{Y}^{(t)}; \tau_1, \bm{\beta}, \eta, \sigma^2)}{\eta+\sigma^2}\left(\frac{\bm{\beta}}{\eta} + \frac{\hat{\bm{\beta}}^{(t)}}{\sigma^2}\right)%=\bar{\bm{\beta}}^{(t)} (\bm{Y}^{(t)})
\end{eqnarray}
as an estimator for $\bm{\beta}^{(t)}$.
%by Proposition \ref{pm}. 
We refer to the $\bar{\bm{\beta}}^{(t)} (\bm{Y})$ as an oracle estimator.
%We want to show that 
Let 
%\begin{equation*}
$\varphi(\alpha) = \int_{\bm{x} \notin \mathcal{B}(\bm{0}, \alpha)} \psi(\bm{x}; \bm{0}, \bm{I}_{p}) d\bm{x},$
%\end{equation*}
where $\mathcal{B}(\bm{0}, \alpha)$ represents the ball centered at $\bm{0}$ with  radius $\alpha$. We provide the theoretical results for the oracle estimator in the following theorem. %, where we assume that the prior parameters are known.

%{\color{red}
\begin{thm}\label{Aoptimal}
	 If $\tau_1$, $\eta$, $\sigma$, and $\bm{\beta}$ are known, then the oracle estimator $\bar{\bm{\beta}}^{(t)} (\bm{Y})$ in Equation (\ref{oracle}) is optimal, that is,
	\begin{equation*}
	R_m(\bar{\bm{\beta}}^{(t)}) = \inf_{\bm{\delta}_m \in \mathcal{E}_m^*} R_m(\bm{\delta}_m) %\to 0
	\end{equation*}
	for each $1\le t\le m$, where $\mathcal{E}^*_m$ is the set consisting of all available estimators of $\bm{\beta}^{(t)}$ with known $\tau_1$, $\eta$, $\sigma$, and $\bm{\beta}$. In addition, for each $1\le t\le m$, 
	%there exists a positive constant $c>0$ such that 
%	{\color{orange}
	\begin{eqnarray}\label{OLScompare}
	%R_m(\hat{\bm{\beta}}^{(t)})-R_m(\bar{\bm{\beta}}^{(t)})>c.
	 R_m(\hat{\bm{\beta}}^{(t)})-R_m(\bar{\bm{\beta}}^{(t)})&\ge& \frac{\sigma^2 \alpha^2  \lambda_{\min} (\Delta)}{\eta+\sigma^2} \left( \tau_1 \varphi \left[ \lambda_{\max}\{(\bm{X}^T\bm{X})^{1/2}\}(2\|\bm{\beta}\|_2+\alpha)/(\sigma^2+\eta)^{1/2}\right] \right. \notag \\
	%\frac{\sigma^2 \alpha^2 (1-\tau_1)}{\eta+\sigma^2} 
	&& \left. + \tau_0 \varphi \left[ \lambda_{\max}\{(\bm{X}^T\bm{X})^{1/2}\}(\|\bm{\beta}\|_2+\alpha)/\sigma \right] \right),
	\end{eqnarray}
%}
	%for sufficiently large $m$.
	where $\alpha$ is any positive constant, and $\lambda_{\min}(\cdot)$ and $\lambda_{\max}(\cdot)$ represent the largest and smallest eigenvalues, respectively.
\end{thm}

Theorem \ref{Aoptimal} states that the oracle estimator $\bar{\bm{\beta}}^{(t)} (\bm{Y})$ can achieve the minimum Bayes risk
%show the optimality of the oracle estimator $\bar{\bm{\beta}}^{(t)} (\bm{Y})$ is 
among all estimators based on known prior parameters.
The equation (\ref{OLScompare}) implies that the oracle estimator is strictly better than the OLS estimator in terms of the Bayes risk function.
In the following theorem, we show the convergence of the proposed estimator to the oracle estimator as the total number of tissues goes to infinity.

\begin{thm}\label{ConsistentInProb}
The proposed estimator $\widetilde{\bm{\beta}}^{(t)} (\bm{Y})$ converges in probability to the oracle estimator, that is,
\begin{equation*}
\left\|\widetilde{\bm{\beta}}^{(t)} (\bm{Y})-\bar{\bm{\beta}}^{(t)} (\bm{Y})\right\|_2 \overset{p}{\to} 0,
\end{equation*}
as $m\to\infty$.
\end{thm}

As shown in Theorem \ref{ConsistentInProb}, when we have more tissues, the proposed estimator gets closer to the optimal oracle estimator $\bar{\bm{\beta}}^{(t)} (\bm{Y})$.
In contrast, the OLS estimator $\hat{\bm{\beta}}^{(t)}$ stays apart from $\bar{\bm{\beta}}^{(t)} (\bm{Y})$. 
This is  due to that the proposed estimator borrows cross-tissue information in the estimation of the common prior parameters, while the OLS estimator only uses information in one single tissue.
%The $\hat{\bm{\beta}}^{(t)}$ and $\bar{\bm{\beta}}^{(t)} (\bm{Y})=\bar{\bm{\beta}}^{(t)} (\bm{Y}^{(t)})$ are both independent of $m$.

%}

%\begin{prop}\label{compare}
%	There exist a positive constant $c>0$ such that $R_m(\hat{\bm{\beta}}^{(t)})-R_m(\widetilde{\bm{\beta}}^{(t)})>c$
%	for sufficiently large $m$.
%\end{prop}

\section{An empirical Bayes regression model with missing data}\label{sec: EB_missing}

In this section, we consider situations where there are missing values in gene expression matrix $\bm{Y}$, which is motivated by missing tissue samples in the GTEx data. Specifically, for each $1\le t\le m$, let $\bm{W}^{(t)}$ be a $n\times n$ diagonal matrix with binary diagonal elements $w_{ii}^{(t)}$ for $1\le i \le n$, $w_{ii}^{(t)}=1$ if and only if $y_{i}^{(t)}$ is observed, where $y_{i}^{(t)}$ is the $i$-th element in $\bm{Y}^{(t)}$.  We assume that each $w_{ii}^{(t)}$  is independent of each other and missing is at random. 
Then, the OLS estimator for $\bm{\beta}^{(t)}$ with missing data is 
\begin{eqnarray}\label{OLS_obs}
\hat{\bm{\beta}}^{(t)}_{obs}=(\bm{X}^T \bm{W}^{(t)} \bm{X})^{-1}\bm{X}^T \bm{W}^{(t)} \bm{Y}^{(t)}=\bm{\beta}^{(t)}+(\bm{X}^T \bm{W}^{(t)} \bm{X})^{-1}\bm{X}^T \bm{W}^{(t)} \bm{\varepsilon}^{(t)}.
\end{eqnarray}
Compared with the OLS estimator in (\ref{OLS}), $\hat{\bm{\beta}}^{(t)}_{obs}$ is constructed only based on %information in the $t$-th tissue from
subjects whose gene expression levels in the $t$-th tissue type are observed.
We also have $\hat{\bm{\beta}}^{(t)}_{obs} \mid\bm{\beta}^{(t)}, \bm{W}^{(t)} \sim N_{p} (\bm{\beta}^{(t)}, \sigma^2(\bm{X}^T \bm{W}^{(t)} \bm{X})^{-1})$.

 Following a similar derivation, we can derive the posterior mean of $\bm{\beta}^{(t)}$ as 
	\begin{eqnarray}\label{postmean_m}
	E(\bm{\beta}^{(t)} \mid  \bm{Y}_{\text{obs}}, \bm{W})
	&=&E(\bm{\beta}^{(t)} \mid  \bm{Y}^{(t)}_{\text{obs}}, \bm{W}^{(t)}) \\
	&=&h_2(\bm{Y}^{(t)}_{\text{obs}}, \bm{W}^{(t)}; \tau_1, \bm{\beta}, \eta, \sigma^2) (\bm{X}^T\bm{X}/\eta+\bm{X}^T \bm{W}^{(t)} \bm{X}/\sigma^2)^{-1} \notag\\
	&& \cdot (\bm{X}^T\bm{X} \bm{\beta}/\eta + \bm{X}^T \bm{W}^{(t)}\bm{Y}^{(t)}/\sigma^2). \notag
\end{eqnarray}
where 
	\begin{equation*}
	h_2(\bm{Y}^{(t)}_{\text{obs}}, \bm{W}^{(t)}; \tau_1, \bm{\beta}, \eta, \sigma^2)=\frac{\tau_1 \psi(\bm{Y}^{(t)}_{\text{obs}}; \bm{X}_t \bm{\beta}, \sigma^2 \bm{I}_{n_t}+ \eta\bm{H}_{t})}{\tau_1 \psi(\bm{Y}^{(t)}_{\text{obs}}; \bm{X}_t \bm{\beta}, \sigma^2 \bm{I}_{n_t}+ \eta\bm{H}_{t})+\tau_0 \psi(\bm{Y}^{(t)}_{\text{obs}}; \bm{0}, \sigma^2 \bm{I}_{n_t})}.
\end{equation*}
Compared with the posterior mean in (\ref{postmean}), the posterior mean of $\bm{\beta}^{(t)}$ in (\ref{postmean_m}) is more complicated since $\bm{X}^T\bm{X}$ and $\bm{X}^T \bm{W}^{(t)} \bm{X}$ are not exactly the same due to missing values. However, $E(\bm{\beta}^{(t)} \mid  \bm{Y}_{\text{obs}}, \bm{W})$ is still a weighted combination of the shared information in $\bm{\beta}$ and the observed information for the $t$-th tissue.
We adopt the posterior expectation with MLEs of $\bm{\beta}, \tau_1, \eta, \sigma^2$
as our proposed estimator. 
To find the MLEs, we provide 
%estimate the parameters $\bm{\beta}, \tau_1, \eta, \sigma^2$ via 
an EM algorithm in Appendix \ref{A_algorithm}, where the estimation for $\bm{\beta}$ is calculated based on observed samples in all the tissues. Under the setting with  missing data, a  Bayes factor and posterior odds ratio can be derived (see Supplemental Materials).

For each tissue $t$, let $\widetilde{\bm{\beta}}^{(t)}_{obs} (\bm{Y}_{\text{obs}}, \bm{W})$ be the   posterior mean given in \eqref{postmean_m} with the parameters estimated using the MLEs,
$\widetilde{\tau}_1$, $\widetilde{\eta}$, $\widetilde{\sigma}$, $\widetilde{\bm{\beta}}$. We define
the Bayes risk function of any estimator $\bm{\delta}_m(\bm{Y}_{\text{obs}}, \bm{W}) \in \bar{\mathcal{E}}_m$ for $\bm{\beta}^{(t)}$ as
%\begin{equation}
%R_m(\bm{\delta}_m)=\int l\left(\bm{\beta}^{(t)}, \bm{\delta}_m(\hat{\bm{B}}^{(m)})\right) \prod_{i=1}^{m} \left\{ p(\hat{\bm{\beta}}^{(i)} \mid \bm{\beta}^{(i)}) p(\bm{\beta}^{(i)}) d\hat{\bm{\beta}}^{(i)} d\bm{\beta}^{(i)} \right\},
%\end{equation}
%{\color{red}
	\begin{equation*}
	R_m^{(obs)}(\bm{\delta}_m)=\int l\left(\bm{\beta}^{(t)}, \bm{\delta}_m(\bm{Y}_{\text{obs}}, \bm{W})\right) \prod_{i=1}^{m} \left\{ p(\bm{W}^{(i)}, \bm{Y}^{(i)}_{\text{obs}} \mid \bm{\beta}^{(i)}) p(\bm{\beta}^{(i)}) d\bm{Y}^{(i)}_{\text{obs}} d\bm{W}^{(i)} d\bm{\beta}^{(i)} \right\},
	\end{equation*}
%}
where $\bar{\mathcal{E}}_m$ is the set consisting of all available estimators of $\bm{\beta}^{(t)}$ with missing data.
When $\eta$, $\sigma$, and $\bm{\beta}$ are known,  let 
$\bar{\bm{\beta}}^{(t)}_{obs} (\bm{Y}_{\text{obs}}, \bm{W})=E(\bm{\beta}^{(t)} \mid \bm{Y}_{\text{obs}}, \bm{W})$ be 
the oracle estimator.  
We demonstrate that the proposed estimator $\widetilde{\bm{\beta}}^{(t)}_{obs} (\bm{Y}_{\text{obs}}, \bm{W})$ is strictly better than the OLS estimator $\hat{\bm{\beta}}^{(t)}_{obs}$ in equation (\ref{OLS_obs}) in terms of the Bayes risk function via the above oracle estimator in the following theorems.

%We want to show that 

%{\color{orange}
%Let 
%\begin{equation}
%\varphi(\alpha) = \int_{\bm{x} \notin \mathcal{B}(\bm{0}, \alpha)} \psi(\bm{x}; \bm{0}, \bm{I}_{p}) d\bm{x},
%\end{equation}
%where $\mathcal{B}(\bm{0}, \alpha)$ represents the ball centered at $\bm{0}$ with  radius $\alpha$.
%}

%{\color{red}
\begin{thm}\label{Aoptimal_m}
	If $\eta$, $\sigma$, and $\bm{\beta}$ are known, the oracle estimator $\bar{\bm{\beta}}^{(t)}_{obs} (\bm{Y}_{\text{obs}}, \bm{W})$ is optimal, that is,
	\begin{equation*}
	R_m^{(obs)}(\bar{\bm{\beta}}^{(t)}_{obs}) = \inf_{\bm{\delta}_m \in \bar{\mathcal{E}}_m^*} R_m^{(obs)}(\bm{\delta}_m) %\to 0
	\end{equation*}
	for each $1\le t\le m$, where $\bar{\mathcal{E}}^*_m$ is the set consisting of all available estimators of $\bm{\beta}^{(t)}$ with known $\eta$, $\sigma$, and $\bm{\beta}$. In addition, for each $1\le t\le m$, 
	%there exists a positive constant $c>0$ such that 
	\begin{eqnarray}\label{OLScompare_m}
	&&R_m^{(obs)}(\hat{\bm{\beta}}^{(t)}_{obs})-R_m^{(obs)}(\bar{\bm{\beta}}^{(t)}_{obs}) \\
	&\ge& \kappa \frac{\sigma^2 \alpha^2  \lambda_{\min} (\Delta)}{\eta+\sigma^2} \left( \tau_1 \varphi \left[ \lambda_{\max}\{(\bm{X}^T\bm{X})^{1/2}\}((1+\kappa)\|\bm{\beta}\|_2+\alpha)/(\sigma^2+\eta)^{1/2}\right] \right. \notag \\
	%\frac{\sigma^2 \alpha^2 (1-\tau_1)}{\eta+\sigma^2} 
	&& \left. + \tau_0 \varphi \left[ \lambda_{\max}\{(\bm{X}^T\bm{X})^{1/2}\}(\kappa\|\bm{\beta}\|_2+\alpha)/\sigma \right] \right), \notag
	\end{eqnarray}
	%for sufficiently large $m$.
	where $\alpha$ is any positive constant 
%	$\lambda_{\min}(\cdot)$ and $\lambda_{\max}(\cdot)$ represent the largest and smallest eigenvalues, respectively, 
	and $\kappa=\lambda_{\min}(\bm{X}^T\bm{X})/\lambda_{\max}(\bm{X}^T\bm{X})$.
\end{thm}
The equation (\ref{OLScompare_m}) in Theorem \ref{Aoptimal_m} shows that the oracle estimator $\bar{\bm{\beta}}^{(t)}_{obs}$ is optimal and has lower Bayes risk 
%is strictly better 
than the OLS estimator $\hat{\bm{\beta}}^{(t)}_{obs}$ for large sample size $n$.
We can similarly show that 
	the proposed estimator $\widetilde{\bm{\beta}}^{(t)}_{obs} (\bm{Y}_{\text{obs}}, \bm{W})$ converges in probability to the oracle estimator, $
	\left\|\widetilde{\bm{\beta}}^{(t)}_{obs} (\bm{Y}_{\text{obs}}, \bm{W}) - \bar{\bm{\beta}}^{(t)}_{obs} (\bm{Y}_{\text{obs}}, \bm{W})\right\|_2 \overset{p}{\to} 0,
$
	as $m\to\infty$.

\section{Simulation} \label{sec: sim}

In this section, we compare the proposed method with the OLS method for handling multi-tissue data. The simulation results show that the proposed method achieves more accurate parameter estimation than the traditional OLS. In each simulation setting,  $100$ replications are performed. For each replication,  
%we generate rows of $\bm{X}$ independently from $N_{p}(\bm{0}, \bm{C})$, where $\bm{C}$ is an exchangeable covariance matrix with all diagonals $1$ and off-diagonals $\rho$. 
we let
\begin{equation*}
\bm{Y}^{(t)}=\bm{X}\bm{\beta}^{(t)}+\bm{\varepsilon}^{(t)},
\end{equation*}
where $\bm{\varepsilon}^{(t)}\sim N_n(\bm{0}, \sigma^2\bm{I}_n)$, 
each row of the $n \times p$ matrix $\bm{X}$ is independent and identically distributed, 
and $\bm{\beta}^{(t)}$ independently follows from a mixture distribution depending on $I^{(t)} \sim B(1, \tau_1)$
for each $t=1, \dots, m$.  

To evaluate the performance of each method, we calculate the mean squared error (MSE) of each estimator $\hat{\bm{\beta}}^{(t)} (1\le t \le m)$: 
\begin{equation*}
\frac{1}{pm}\sum_{t=1}^m\|\hat{\bm{\beta}}^{(t)}-\bm{\beta}^{(t)}\|_2^2,
\end{equation*} 
which measures the parameter estimation accuracy across all tissues. We say that a method has better performance if the MSE is smaller. We compare the proposed method and the OLS under the following settings.  In addition,  we use the posterior probability of $I^{(t)}$ produced by the proposed method to calculate the area under the receiver operating characteristic curve (AUC) for identifying the tissues with non-zero cis-effects. Larger AUC indicates better performance of the posterior probabilities on prediction of $I^{(t)}$.
In the first two settings, we assume that there is no missing values in the response $\bm{Y}^{(t)}$. The difference of the two settings mainly comes from the generation of $\bm{\beta}^{(t)}$. In contrast, we consider the responses with random missing values in Settings \ref{setting_5} and \ref{setting_6}, where we adopt the proposed estimator in Section \ref{sec: EB_missing} to deal with the missing responses. Moreover, in Setting \ref{setting_6}, we directly use the cis-SNPs in the real GTEx genotype data as covariates $\bm{X}$ to capture in real linkage disequilibrium structures of the genotype data.

\begin{setting}\label{setting_3}
%\textbf{Setting $3$}: 
Let $p=30$, $n=50$, $m=50$, $\tau_1=0.5$, $\bm{\beta}=(\beta_s \bm{1}_{10}^T, (\beta_s/2) \bm{1}_{10}^T, \bm{0}_{10}^T)^T$,  and $\sigma^2=10$, where $\beta_s$ is the signal level. We generate each row of $\bm{X}$ from $N_{p}(\bm{0}, \bm{C})$
and generate $\bm{\beta}^{(t)}$ from
 \begin{eqnarray*}
	&&\bm{\beta}^{(t)} \mid I^{(t)}=1 \sim N_{p} (\bm{\beta}, \bm{C}),\\
	&&\bm{\beta}^{(t)} \mid I^{(t)}=0 \equiv \bm{0},
\end{eqnarray*}
for $t=1, \dots, m$, 
where $\bm{C}$ is an exchangeable covariance matrix with all diagonals $1$ and off-diagonals $\rho$. 
%We follow similarly as in Setting \ref{setting_1},  except that $m=50$ and $\sigma^2=10$.
\end{setting}
%Sim16C.R

In Setting \ref{setting_3}, we generate $\bm{\beta}^{(t)}$ following our model construction in Section \ref{sec: model}, but we do not set the covariance matrix of $\bm{\beta}^{(t)}$ to be exactly $\eta(\bm{X}^T\bm{X})^{-1}$. Nevertheless, as shown in Table \ref{Settings34} the proposed method still performs better than the OLS under various correlation and signal levels. For example, when $\rho=0.6$ and $\beta_s=2$, 
the MSE of the OLS is $13.7753$, 
while 
the MSE of the proposed method is $2.0096$
which is only $15\%$ of that of the OLS.  
Moreover, AUCs in Table \ref{Settings34} are all higher than $0.6$, especially for situations with large signal levels, indicating the the propose method can identify the tissues with non-zero cis-SNPs for a given gene. 
Under Setting \ref{setting_3}, we let $\tau_1=0.5$, indicating that the two groups of tissues ($\bm{\beta}^{(t)}\neq \bm{0}$ verse $\bm{\beta}^{(t)}=\bm{0}$) are balanced. In Appendix \ref{A_simulation}, we provide simulations with different $\tau_1$ where the two groups of tissues are unbalanced. The proposed method still outperforms OLS in terms of MSE.
%As shown in Table \ref{Settings34}, although Setting \ref{setting_3} has a larger error term,
%%and $4$ are more challenging than Settings $1$ and $2$, 
%the proposed method still performs much better than the traditional OLS. For example,  when $\rho=0.6$ and $\beta_s=2$, the MSE of the proposed method is only $15\%$ 
%%and $7\%$ 
%of that of the OLS.
%% in Settings $3$ and $4$, respectively. 

\begin{table}[H]
	\centering
	\caption{MSEs of OLS and the proposed method, and AUCs of the propsoed method for identifying tissues with non-zero cis-SNPs under Settings \ref{setting_3} and \ref{setting_4}.}
	\resizebox{\textwidth}{!}{
	\begin{tabular}{c|c|ccc|ccc}
		\hline
		\hline
	 \multirow{2}{*}{Correlation $\rho$} & \multirow{2}{*}{Signal level $\beta_s$} & \multicolumn{3}{c|}{Setting \ref{setting_3}} &  \multicolumn{3}{c}{Setting \ref{setting_4}}\\ 
	\cline{3-8}
	  &  & OLS & Proposed & AUC & OLS & Proposed &AUC \\ 
  \hline
 0 & 0.5 & 6.6831 & 0.7442 & 0.8877 & 0.0597 & 0.0166 & 0.8467 \\ 
 %0 & 0.8 & 5.3746 & 0.4844 & 0.8851 & 0.0600 & 0.0084 & 0.5216 \\ 
 0 & 1.0 & 5.8465 & 0.7509 & 0.9696 & 0.0491 & 0.0090 & 0.8478 \\ 
 0 & 2.0 & 5.7021 & 0.5427 & 1.0000 & 0.0478 & 0.0086 & 0.7965 \\ 
0.2 & 0.5 & 6.6928 & 0.5362 & 0.6763 & 0.0732 & 0.0099 & 0.5668 \\ 
%0.2 & 0.8 & 5.9603 & 0.6063 & 0.9026 & 0.0647 & 0.0131 & 0.8393 \\ 
0.2 & 1.0 & 5.8547 & 0.6186 & 0.8958 & 0.0624 & 0.0039 & 0.6200 \\ 
0.2 & 2.0 & 5.6966 & 0.6872 & 0.9952 & 0.0654 & 0.0087 & 0.7949 \\ 
 0.4 & 0.5 & 7.6919 & 1.3563 & 0.8942 & 0.0788 & 0.0147 & 0.4913 \\ 
 %0.4 & 0.8 & 7.6807 & 0.5531 & 0.5649 & 0.0988 & 0.0154 & 0.7308 \\ 
 0.4 & 1.0 & 8.5119 & 0.6462 & 0.8301 & 0.0686 & 0.0093 & 0.6571 \\ 
 0.4 & 2.0 & 7.4002 & 1.3132 & 0.9507 & 0.0963 & 0.0064 & 0.8636 \\ 
 0.6 & 0.5 & 11.6754 & 1.0297 & 0.6907 & 0.1254 & 0.0159 & 0.7296 \\ 
 %0.6& 0.8 & 13.1505 & 0.8115 & 0.6783 & 0.1353 & 0.0126 & 0.6638 \\ 
 0.6 & 1.0 & 13.8658 & 2.9334 & 0.8966 & 0.1442 & 0.0126 & 0.8213 \\ 
 0.6& 2.0 & 13.7753 & 2.0096 & 1.0000 & 0.1728 & 0.0121 & 0.8535 \\ 
 0.8 & 0.5 & 26.6533 & 3.8958 & 0.8200 & 0.2575 & 0.0125 & 0.4911 \\ 
 %0.8 & 0.8 & 26.3922 & 5.8356 & 0.8591 & 0.2829 & 0.0125 & 0.5373 \\ 
 0.8 & 1.0 & 28.0534 & 12.0643 & 0.9491 & 0.2492 & 0.0149 & 0.7135 \\ 
 0.8 & 2.0 & 27.4862 & 6.6816 & 0.9279 & 0.2555 & 0.0115 & 0.8558 \\
		\hline
		\hline
	\end{tabular}}
	\label{Settings34}
\end{table}

%In the following setting, we generate $\bm{\beta}^{(t)}$ differently from our model construction in Section \ref{sec: model}.

%\newpage
\begin{setting}\label{setting_4}
%\textbf{Setting $4$}: 
We follow similarly as in Setting \ref{setting_3} except that $\sigma^2=1$ and 
 \begin{eqnarray*}
	&&\beta^{(t)}_{1} \mid I^{(t)}=1 \sim N (\beta_s, \sigma^2),\\
	&&\beta^{(t)}_{1} \mid I^{(t)}=0 \equiv 0,
\end{eqnarray*}
where $\beta^{(t)}_{1}$ is the first element in $\bm{\beta}^{(t)}$, other elements in $\bm{\beta}^{(t)}$ are all zero for $t=1, \dots, T$.
\end{setting}
%Sim18C.R

In Setting \ref{setting_4}, we investigate robustness of the proposed method when $\bm{\beta}^{(t)}$ is not fully generated from our assumed mixture distribution in Section \ref{sec: model}. Under this setting, AUCs are lower than those in Setting \ref{setting_3}, but most of AUCs are still higher than $0.5$. In addition, the proposed method produces much smaller MSE than the OLS. For instance, at $\rho=0.8$ and $\beta_s=2$, the MSE of the proposed method is only $5\%$ 
%and $7\%$ 
of that of the OLS. 
%Note that the AUC in this case is only $0.4911$.
Thus, the proposed method is robust to certain errors in the model assumption.

In real genetic data such as the GTEx data, we may not collect all tissues from each subject, indicating that the gene expression matrix $\bm{Y}$ could contain missing values. 
%Thus, we extend the proposed method in Appendix \ref{A_method} and 
%As mentioned in Section \ref{sec: EB_missing}, we
We provide the proposed estimator in Section \ref{sec: EB_missing}
%an algorithm in Appendix \ref{A_algorithm} 
to handle cases with missing values in $\bm{Y}$. In the following Setting \ref{setting_5}, we investigate the performance of the proposed method when there are  missing data.

\begin{setting}\label{setting_5}
We follow similarly as in Setting \ref{setting_3}
except that,  for each $1\le t \le m$,  we random select $20\%$ elements in $\bm{Y}^{(t)}$ and set them to be missing.
% except that $\sigma^2=1$ and 
% \begin{eqnarray*}
%	&&\beta^{(t)}_{1} \mid I^{(t)}=1 \sim N (\beta_s, \sigma^2),\\
%	&&\beta^{(t)} \mid I^{(t)}=0 \equiv 0,
%\end{eqnarray*}
%where $\beta^{(t)}_{1}$ is the first element in $\bm{\beta}^{(t)}$, other elements in $\bm{\beta}^{(t)}$ are all zero for $t=1, \dots, T$.
\end{setting}
%Sim16CMiss.R

We provide results of Setting \ref{setting_5} in Table \ref{Settings56}. 
Due to missing values, the MSEs of the OLS in Setting \ref{setting_5} are larger than those in Setting \ref{setting_3}.
%Although the MSEs of OLS and the proposed method in Setting \ref{setting_5} are larger than those in Setting \ref{setting_3} due to missing values, 
However, the MSEs of the proposed method in Setting \ref{setting_5} do not change much compared to those in Setting \ref{setting_3},
implying that the proposed method produces much smaller MSE than the OLS. For instance, when $\rho=0$ and $\beta_s=0.5$, the MSE of the proposed method is only $6\%$ of that of OLS.

\begin{setting}\label{setting_6}
We follow similarly as in Setting \ref{setting_5}
except $n=300$ and that,
%and $\sigma^2=1$.
%In addition,  
for the construction of $\bm{X}$,  
we randomly select $300$ samples independently with replacement from the $838$ samples of the $30$ cis-SNPs of the gene \textit{WARS2} in the GTEx data.
%for each $1\le t \le m$,  we random select $20\%$ elements in $\bm{Y}^{(t)}$ and set them to be missing.
% except that $\sigma^2=1$ and 
% \begin{eqnarray*}
%	&&\beta^{(t)}_{1} \mid I^{(t)}=1 \sim N (\beta_s, \sigma^2),\\
%	&&\beta^{(t)} \mid I^{(t)}=0 \equiv 0,
%\end{eqnarray*}
%where $\beta^{(t)}_{1}$ is the first element in $\bm{\beta}^{(t)}$, other elements in $\bm{\beta}^{(t)}$ are all zero for $t=1, \dots, T$.
\end{setting}
%Sim22CMiss_m8.R

As shown in Table \ref{Settings56}, the proposed method outperforms the OLS in terms of MSE when we use the real cis-SNP genotype covariates in GTEx. 
For example, with $\rho=0.4$ and $\beta_s=1$, the MSE of the proposed estimator is $0.8710$, while the MSE of the OLS estimator is $4.7770$ which is five times larger than that of the proposed estimator.
In addition, the AUCs of the proposed method across different correlations and signal levels are at least as high as $82\%$, indicating that the posterior probabilities of $T^{(t)}$ produced by the proposed method can effectively identify tissues with non-zero $\bm{\beta}^{(t)}$.

\begin{table}%[H]
	\centering
	\caption{MSEs of OLS and the proposed method, and AUCs of the propsoed method for identifying tissues with non-zero cis-SNPs under Settings \ref{setting_5} and \ref{setting_6}.}
	\resizebox{\textwidth}{!}{
	\begin{tabular}{c|c|ccc|ccc}
		\hline
		\hline
	 \multirow{2}{*}{Correlation $\rho$} & \multirow{2}{*}{Signal level $\beta_s$} & \multicolumn{3}{c|}{Setting \ref{setting_5}} &  \multicolumn{3}{c}{Setting \ref{setting_6}}\\ 
	\cline{3-8}
	  &  & OLS & Proposed & AUC 
	  & OLS & Proposed &AUC \\ 
  \hline
 0 & 0.5 & 11.1561 & 0.7171 & 0.7920 
 & 4.7869 & 0.5939 & 0.9164\\ 
% 0.0000 & 0.8000 & 11.2413 & 0.6918 & 0.8832 \\ 
 0 & 1.0 & 11.0515 & 0.6659 & 0.9419 
 & 4.8158 & 0.6461 & 0.9986\\ 
 0 & 2.0 & 10.9929 & 0.6511 & 0.9974 
 & 4.7832 & 0.6829 & 1.0000\\ 
 0.2 & 0.5 & 13.2371 & 0.6223 & 0.7099 
 & 4.7404 & 0.7378 & 0.8256\\ 
% 0.2000 & 0.8000 & 13.6755 & 0.6336 & 0.8187 \\ 
 0.2 & 1.0 & 13.4058 & 0.6738 & 0.8895 
 & 4.7819 & 0.7492 & 0.9365\\ 
 0.2 & 2.0 & 13.7634 & 0.8114 & 0.9838 
 & 4.8236 & 0.9807 & 0.9880\\ 
 0.4 & 0.5 & 17.5384 & 1.5099 & 0.8031 
 & 4.7390 & 0.9790 & 0.8919\\ 
% 0.4000 & 0.8000 & 17.9321 & 1.2141 & 0.7928 \\ 
  0.4 & 1.0 & 18.1750 & 1.0159 & 0.8420 
  & 4.7770 & 0.8710 & 0.9068\\ 
 0.4 & 2.0 & 18.4120 & 1.3680 & 0.9480 
 & 4.7679 & 1.0975 & 0.9551\\ 
 0.6 & 0.5 & 26.8559 & 3.9957 & 0.8747 
 & 4.7603 & 1.1269 & 0.9348\\ 
% 0.6000 & 0.8000 & 27.0125 & 2.9391 & 0.8338 \\ 
 0.6 & 1.0 & 26.5114 & 2.8599 & 0.8385 
 & 4.7758 & 1.0082 & 0.9067\\ 
 0.6 & 2.0 & 26.6272 & 2.9954 & 0.9074 
 & 4.7527 & 1.1255 & 0.9346\\ 
0.8 & 0.5 & 53.7184 & 10.9895 & 0.9120 
& 4.7581 & 1.1468 & 0.9263\\ 
%  0.8000 & 0.8000 & 53.2167 & 9.8077 & 0.8896 \\ 
 0.8 & 1.0 & 54.1613 & 9.6208 & 0.8912 
 & 4.7885 & 1.0874 & 0.9049\\ 
   0.8 & 2.0 & 54.2999 & 8.6471 & 0.8955 
   & 4.7350 & 1.1774 & 0.9284\\ 
		\hline
		\hline
	\end{tabular}}
	\label{Settings56}
\end{table}

%{\color{red}
\section{Application to GTEx data} \label{sec: real}
%}

In this section, we apply the proposed method to the Genotype-Tissue Expression (GTEx) data \citep{gtex} and compare it with the traditional OLS method in term of predicting tissue-specific expressions using the cis-SNP data. We are also interested in understanding the cis-SNP and gene expression associations across different issue. The GTEx is an ongoing US National Institutes of Health (NIH) Common Fund project starting from 2010, which aims to establish a comprehensive public resource database for investigation of the relationship between genetic variation and gene expression. The GTEx project collects non-diseased tissue samples from nearly $1000$ donors, which are sent to the Laboratory, Data Analysis and Coordinating Center (LDACC) for molecular analysis \citep{lonsdale2013genotype}.
DNA from each donor's blood sample is genotyped using the Illumina HumanOmni5M-Quad BeadChip for whole-genome SNP \citep{lonsdale2013genotype}, and the Illumina TrueSeq RNA sequencing is used for the measurement of gene expression.

Specifically, the GTEx project includes genotype and gene expression data for $838$ participants across $49$ tissue types. 
%includes publicly available genotype, gene expression
%The GTEx data contains gene expressions across $49$ tissue types and genotype data from $838$ participants. 
There are $8066$ genes with expression data available in all the tissue types.
%However, for each tissue type, 
%there are participants whose corresponding 
Since tissue samples of some participants are not collected for each tissue type, 
our analyses of GTEx data focus on $32$ tissues,  each of which has at least $n=200$ collected samples.

We apply the standard data pre-processing and quality-control procedures to both DNA and RNA sequencing data. We exclude the SNPs with minor allele frequencies less than $5\%$. The SNPs are further pruned for linkage disequilibrium (LD)
with a window size of 50 SNPs, a step size of 5 SNPs, and a R2 threshold of 0.2 using PLINK 1.9.
In addition, we select cis-eQTLs for each gene following \cite{wang2016imputing}, which can be viewed as a screening of predictors. Specifically, we first obtain tissue-specific cis-eQTLs of pairs of genes and its corresponding cis-SNPs using the ``MatrixEQTL'' R package (\url{https://cran.r-project.org/web/packages/MatrixEQTL/index.html}
%{\ttfamily https://cran.r-project.org/web/\linebreak packages/MatrixEQTL/index.html}
). For each pair, we then combine the Z statistics of the cis-eQTLs from all the tissues via the Stouffer's Method \citep{stouffer1949american}. We use the ``poolr'' R package (\url{https://cran.r-project.org/web/packages/poolr/index.html}) to carry out the Stouffer's Method. We select the cis-eQTLs whose Stouffer's p values are less than $10^{-6}$. Then there are $4827$ genes with at least one selected cis-eQTL.
%, which accordingly have at least one selected cis-SNP. 
For each gene, we order the selected cis-SNPs by the corresponding Stouffer's p values increasingly.
To avoid highly correlated cis-SNPs, we retain the selected cis-SNPs in a increasing order of the corresponding Stouffer's p values, and remove cis-SNPs which are highly correlated with previously retained cis-SNPs with the coefficient of determination larger than $0.5$.

\subsection{Prediction and comparison with OLS}
We apply the proposed method and the OLS to each gene and its corresponding cis-SNPs. To evaluate each method in gene expression prediction using cis-SNPs, we use a $10$-fold cross-validation analysis. Specifically, we randomly split the observed samples in each tissue into $10$ equally sized subsamples, named from Subsample $1$ to Subsample $10$. For each $1\le i \le 10$, we use Subsample $i$ in all the tissues as a testing set, and the remaining $9$ subsamples in all the tissues as a training set. 
%On each testing set, 
We predict the gene expression values of subjects in each testing set for each gene based on each method. For the proposed method, we use the extension version in Section \ref{sec: EB_missing} that  can handle missing values, since there are missing samples in some tissues for a subject in the training set.
We repeat this procedure $10$ times and obtain predicted values for all the subjects. 
To evaluate the prediction accuracy of each method, we calculate the prediction mean squared error (PMSE) and squared Pearson correlation ($R^2$) between the predicted values and true gene expression values for each gene and each tissue type.

\begin{figure}%[H]     % /Users/feixue/Dropbox/Tensor GTEx/Empirical Bayes/Real data/EQTL_correct/Cor_eQTL/Cor_remove2_cluster/SXY_combine1
	%\captionsetup{font=small}
	\begin{center}
		\resizebox{430.632pt}{420pt}{\includegraphics{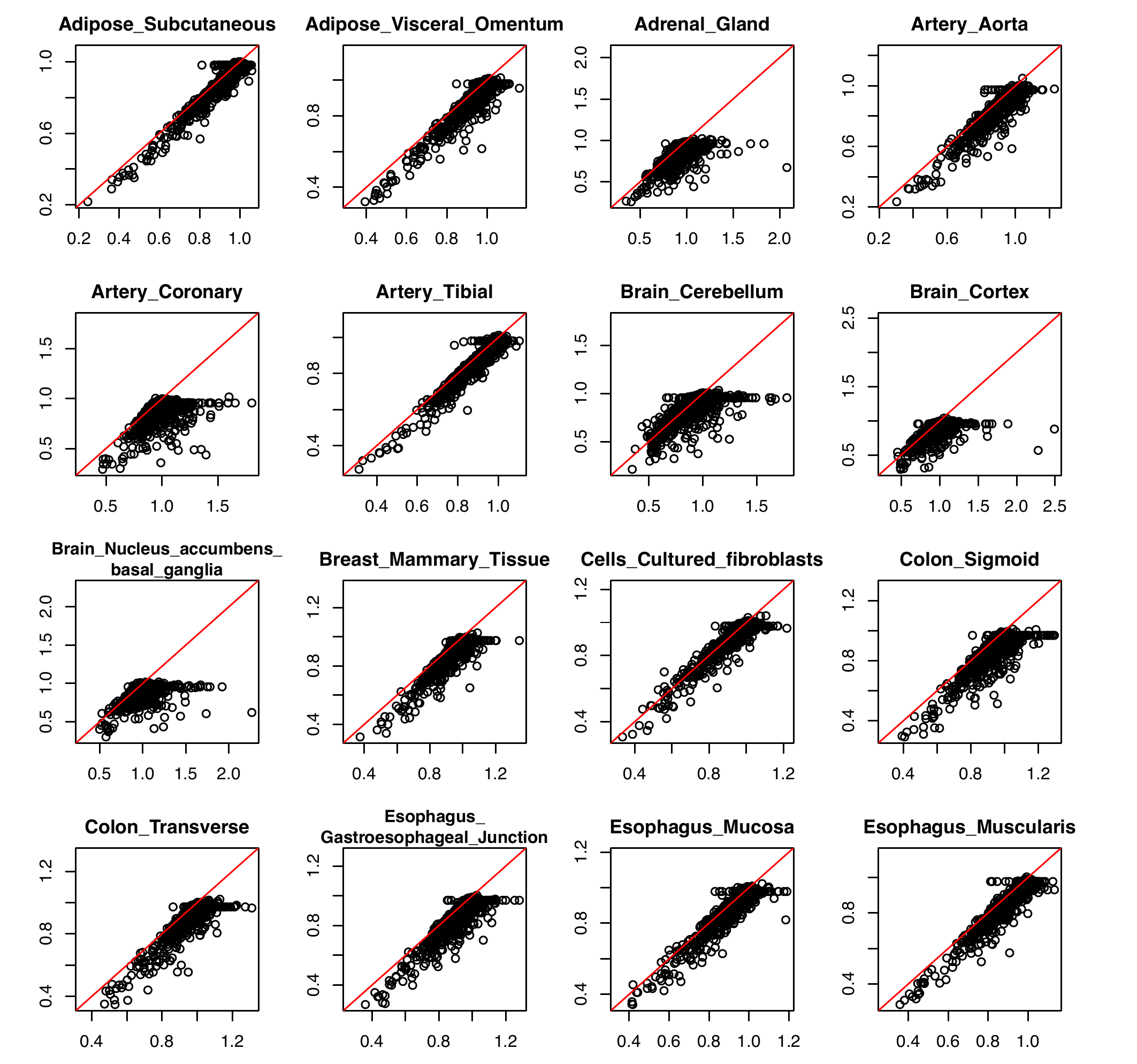}}
		%		\resizebox{160.632pt}{207pt}{\includegraphics{SPAC1.png}} %\hfill 
		%		\hspace{15mm}\resizebox{160.632pt}{207pt}{\includegraphics{SPAC2.png}}
		\caption{Scatter plots of PMSEs in various tissues. Y-axis: PMSE of the proposed method. X-axis: PMSE of the OLS.}\label{PMSE_1}
	\end{center}
	%\vspace{-2mm}
\end{figure}

\begin{figure}%[H]  % /Users/feixue/Dropbox/Tensor GTEx/Empirical Bayes/Real data/EQTL_correct/Cor_eQTL/Cor_remove2_cluster/SXY_combine2
	%\captionsetup{font=small}
	\begin{center}
		\resizebox{430.632pt}{420pt}{\includegraphics{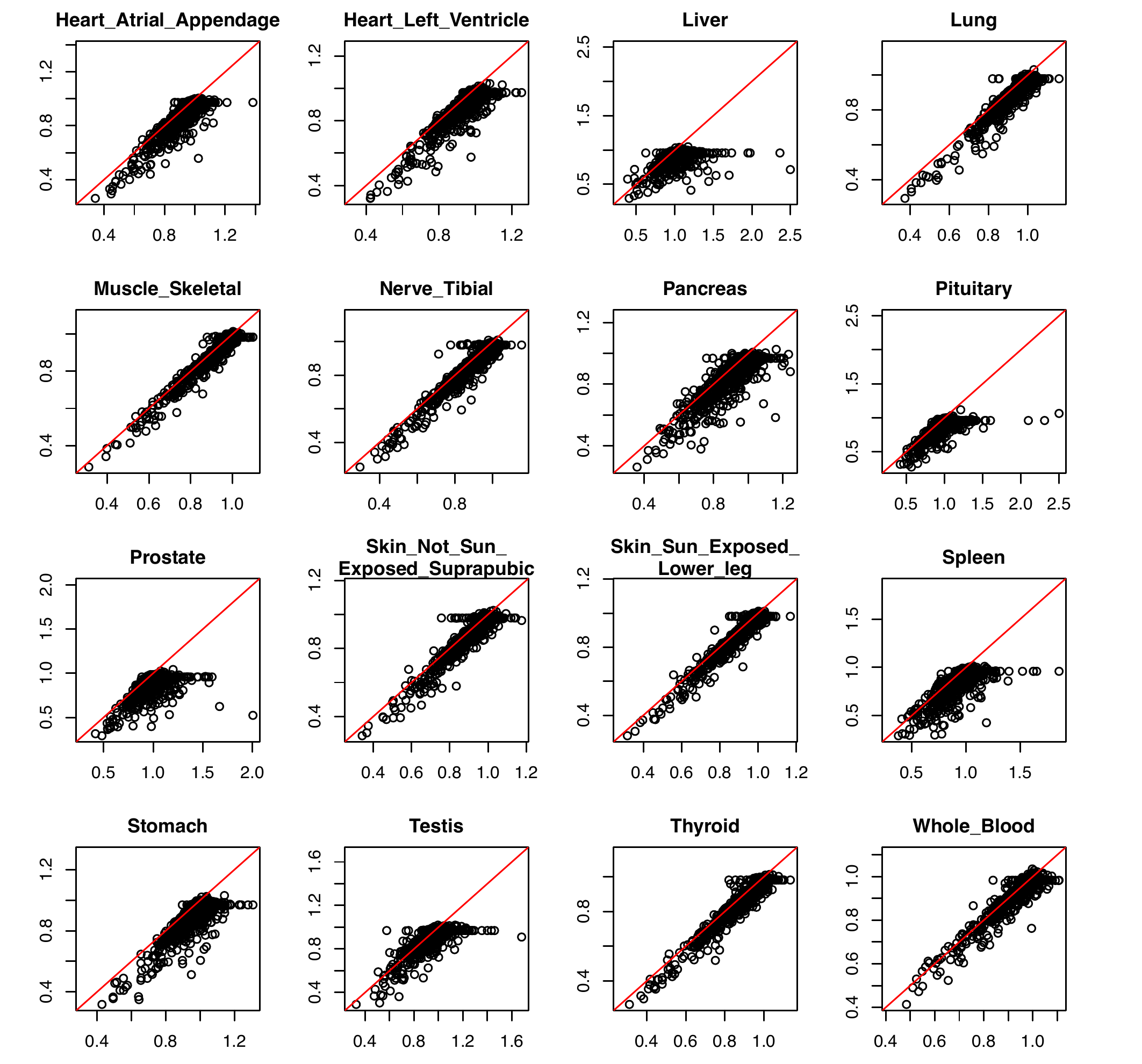}}
		%		\resizebox{160.632pt}{207pt}{\includegraphics{SPAC1.png}} %\hfill 
		%		\hspace{15mm}\resizebox{160.632pt}{207pt}{\includegraphics{SPAC2.png}}
		\caption{Scatter plots of PMSEs in various tissues. Y-axis: PMSE of the proposed method. X-axis: PMSE of the OLS.}\label{PMSE_2}
	\end{center}
	%\vspace{-2mm}
\end{figure}

We %then take average of the PMSEs in all the $10$ testing set, and 
provide the PMSEs of the proposed method and the OLS method for all the genes and tissue types in Figures \ref{PMSE_1} and \ref{PMSE_2}. Each sub-figure in Figures \ref{PMSE_1} and \ref{PMSE_2} is a scatter plot of PMSEs for all genes in one tissue type, where x-axis represents PMSE of the proposed method, and y-axis represents PMSE of the OLS method. In each sub-figure, most points are under the red line where the PMSEs of the two methods are the same. 
Especially, in the brain cortical tissue and the prostate tissue, for some genes, the PMSEs of the proposed method is even smaller than $50\%$ of the corresponding PMSEs of the OLS.
Thus, the proposed method overall performs better than the OLS in terms of prediction accuracy, indicating that the proposed method estimates parameters in (\ref{model}) more accurately than the OLS.

\begin{figure}%[H]  % /Users/feixue/Dropbox/Tensor GTEx/Empirical Bayes/Real data/EQTL_correct/Cor_eQTL/Cor_remove2_cluster
	%\captionsetup{font=small}
	\begin{center}
		\includegraphics[scale=0.6]{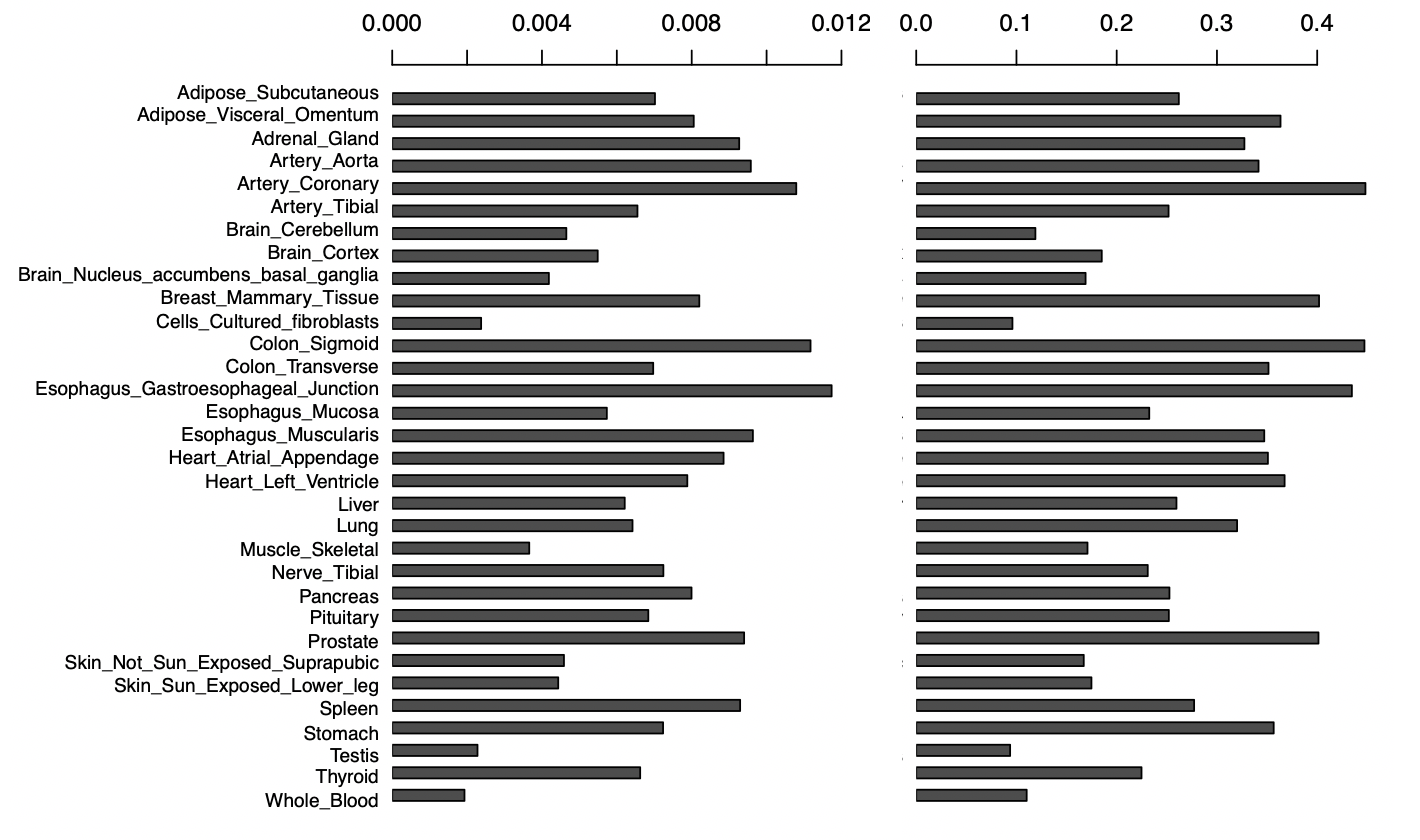}
		\includegraphics[scale=0.6]{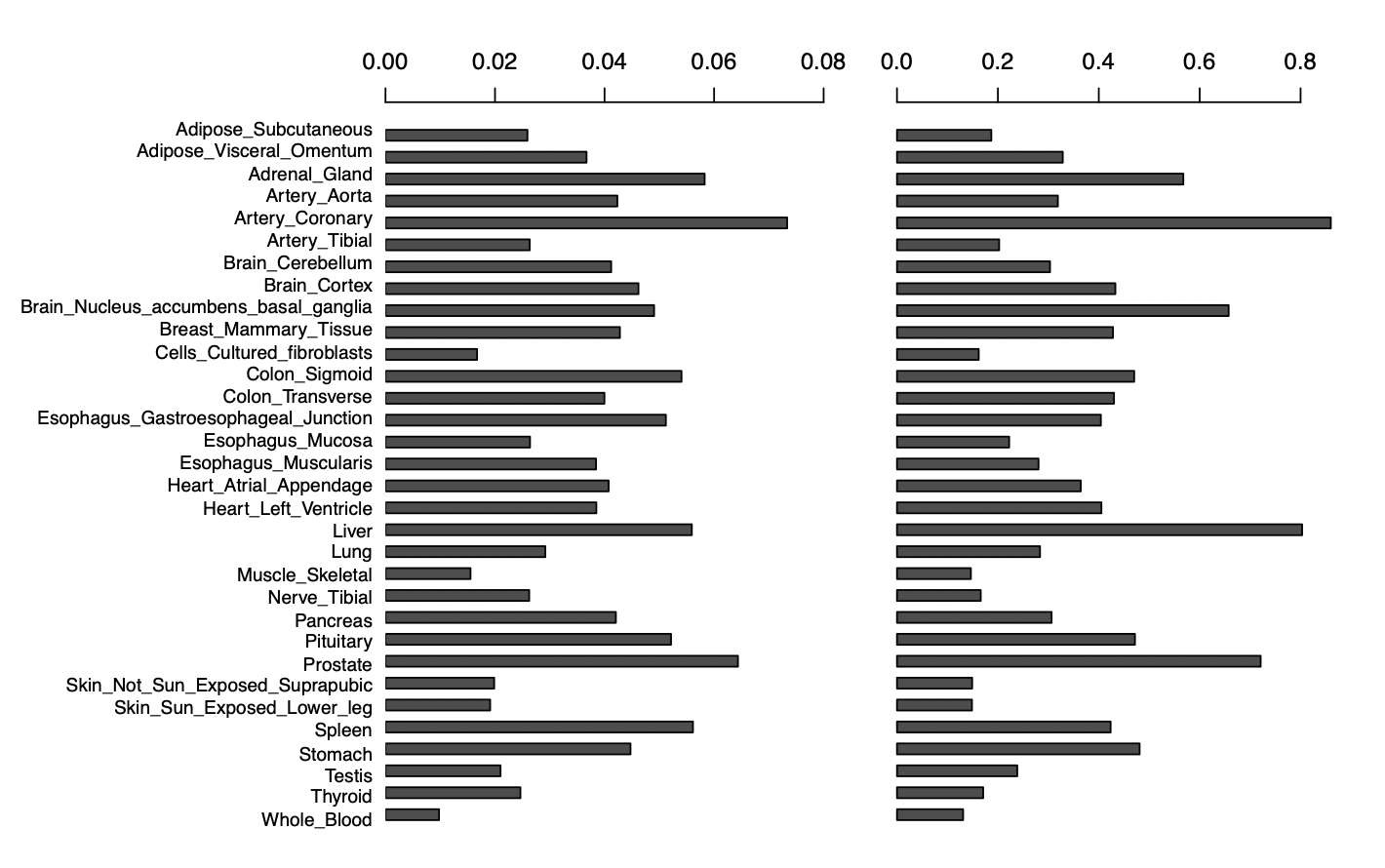}
		%	\includegraphics[scale=0.55]{R_square_increase.pdf}
		%		\includegraphics[scale=0.55]{R_square_increase_relative.pdf}
		%			\includegraphics[scale=0.55]{R_square_increase_20.pdf}
		%		\includegraphics[scale=0.55]{R_square_increase_relative_20.pdf}
		%\resizebox{490.632pt}{350pt}{\includegraphics{R_square_increase_relative.pdf}}
		%		\resizebox{160.632pt}{207pt}{\includegraphics{SPAC1.png}} %\hfill 
		%		\hspace{15mm}\resizebox{160.632pt}{207pt}{\includegraphics{SPAC2.png}}
		%\vspace{-10mm}
		\caption{Increase of $R^2$ by the proposed method compared with the OLS method. Top left: Average increase of $R^2$ by the proposed method. Top right: Percentage of average increase of $R^2$ by the proposed method. Bottom left: Average increase of $R^2$ for genes with at least $20$ cis-SNPs by the proposed method. Bottom right: Percentage of average increase of $R^2$ for genes with at least $20$ cis-SNPs by the proposed method.}\label{Increase}
	\end{center}
	%\vspace{-2mm}
\end{figure}

For each tissue type, we also calculate the average increment and the percentage of the average increment of $R^2$ by the proposed method compared with the OLS method across all the genes. We provide these comparison results for all the tissues in Figure \ref{Increase}. Note that the proposed method substantially increases the average $R^2$ in all the tissues compared to the OLS method, which indicates that the proposed method produces the predicted values more correlated  to the true gene expression values.
On average, the proposed method improves the $R^2$ by $27.5\%$ across all the tissue types.
In particular, for the artery coronary tissue, the proposed method increases the $R^2$ by more than $40\%$.
%of the proposed method increases over $40\%$ of that of the OLS method. 
When we only consider genes with at least $20$ cis-SNPs, the percentage of average increase of $R^2$ by the proposed method is over $80\%$ in the artery coronary tissue.

\subsection{Posterior probability of cis-SNP gene expression association}
In this subsection, we apply the proposed method to the whole dataset, and
%Based on the proposed method, we 
calculate the posterior probability of $I^{(t)}=1$ for $t=1, \dots, m$ and all the genes.
This posterior probability indicates whether the cis-SNPs of a given gene have effects on the corresponding gene expression in a particular tissue based on the observed data.
We %provide
use the ``gplots'' R package (\url{https://cran.r-project.org/web/packages/gplots/index.html}) to generate
a heat map for all the posterior probabilities, where each column represents a tissue and each row represents a gene. 
We first note that for many genes, the cis-SNP and gene expression associations are observed all the tissues (top  blue rows). There are only a few genes that do not have their corresponding cis-SNPs in all the tissues (bottom  red rows). For many other genes, we observe tissue-specific cis-SNP gene expression associations, but for most of these genes, the cis-SNP associations are observed in most of the tissues. 

\begin{figure}%[H]  %/Users/feixue/Dropbox/Tensor GTEx/Empirical Bayes/Real data/EQTL_correct/Cor_eQTL/Cor_remove2_w_cluster
	%\captionsetup{font=small}
	\begin{center}
%	\hspace{-15mm}
	\resizebox{430.632pt}{630pt}{\includegraphics{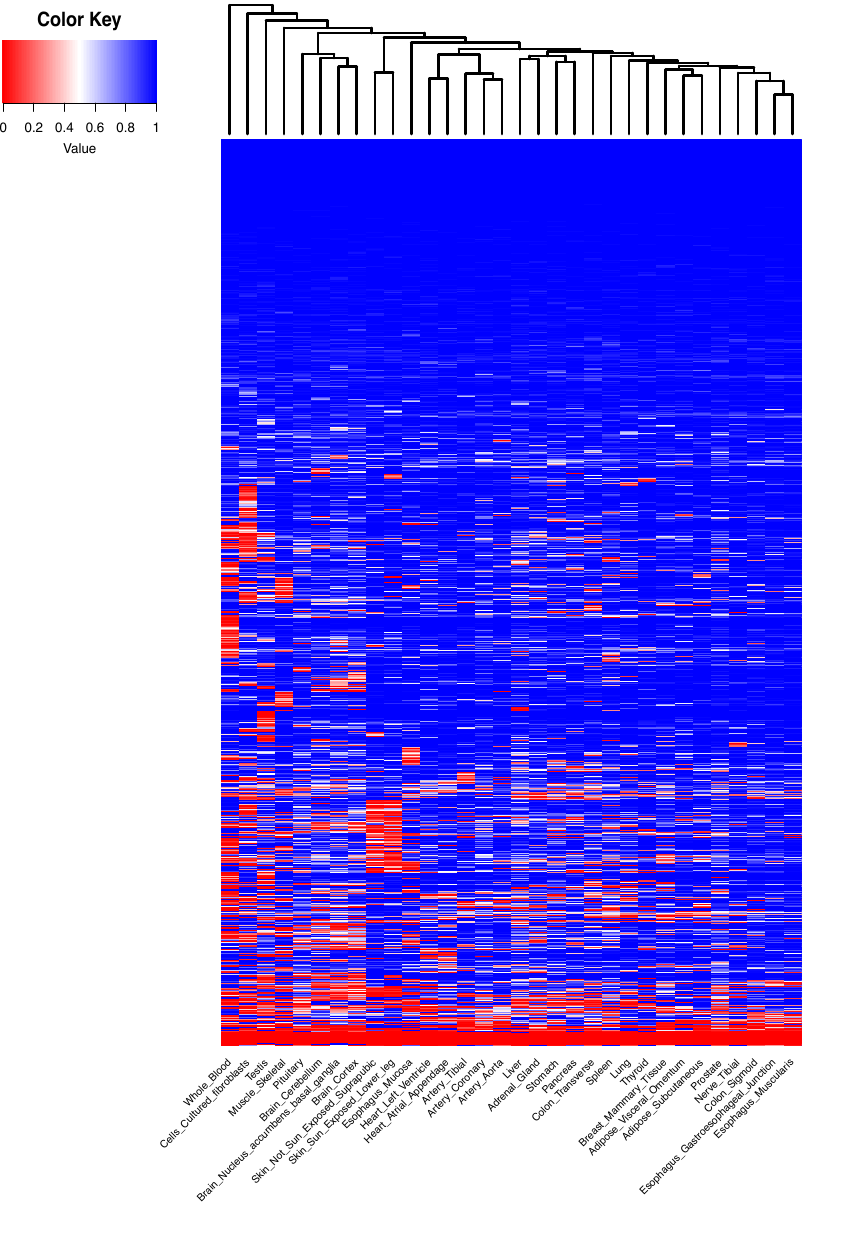}}
%		\resizebox{160.632pt}{207pt}{\includegraphics{SPAC1.png}} %\hfill 
%		\hspace{15mm}\resizebox{160.632pt}{207pt}{\includegraphics{SPAC2.png}}
\vspace{-10mm}
		\caption{Posterior probabilities of $I^{(t)}=1$ for all the genes and tissues.}\label{postprob}
	\end{center}
	%\vspace{-2mm}
\end{figure}

As shown in Figure \ref{postprob}, the heat map clusters similar genes and similar tissues together based on the posterior probability of observing cis-SNP and gene expression association.  We observe  that the tissues that  are clustered together are indeed  For example, the ``Esimilar. sophagus\_Gastroesophageal\_Junction'' tissue and the ``Esophagus\_Muscularis'' tissue are clustered together in Figure \ref{postprob}. They are both related to the esophagus. Similarly, the ``Artery\_Tibial'',
``Artery\_Coronary'', and ``Artery\_Aorta'' tissues are all related to the artery and are clustered together in the heat map. Thus, the posterior probabilities based on the proposed method indeed capture the similarity between tissues in terms of the relationship between gene expression and cis-SNPs.

\section{Discussion}
We develop a new empirical Bayes regression model for multi-tissue eQTL analysis using the GTEx data, which improves the single tissue analysis produced by the traditional ordinary least squares  method.
%to estimate the tissue-specific effects of cis-SNPs on the expression levels of a gene based on the GTEx data. 
To borrow information across tissues, the proposed method assigns a common mixture prior distribution to the cis-SNP effects in each tissue, and estimates parameters in the prior distribution through maximizing marginal likelihood of the expression levels in all the tissues. In addition, the method provides a way of quantify the evidence  whether the cis-SNPs are ``active'' or not in a certain tissue based on the posterior probabilities of the latent configuration indicator in the mixture prior distribution. We apply  the EM algorithm to find the maximum likelihood estimate of prior parameters. Moreover, to accommodate real data with missing responses such as the GTEx data, we have also developed the empirical Bayes estimator and the corresponding asymptotic results when there are  missing data.

Theoretically, we have shown that the proposed estimator is asymptotically superior than the OLS estimator in terms of the Bayes risk. This superiority is mainly due to that the OLS only uses single tissue information while the proposed method incorporates common information from other tissues.
In addition, the application to the GTEx data illustrates that the proposed method estimates the tissue-specific cis-SNP effects more accurately than the OLS method. More importantly, the proposed method provides posterior probabilities of whether there is cis-effects or not for each tissue, which indeed reflects similarity among tissues. For instance, the ``Artery\_Tibial'', ``Artery\_Coronary'', and ``Artery\_Aorta'' tissues are shown to have similar cis-SNP gene expression association among all the genes.

In general, the empirical Bayes method provides  a powerful framework for pooling information across multiple experiments or sources, and improving the accuracy of the estimation or inference in each experiment.
Besides the SNP-gene association, we can also extend the empirical Bayes framework to improve the estimation of the relationship between expression levels of genes for the GTEx project where gene expression levels are measured over multiple tissues. For example, we could incorporate information across tissues through estimating a common prior on the multiple precision matrices for the multiple tissues. 
In addition, in this article, we mainly consider the association between gene expression and cis-SNPs. It would be of great interest to 
%For future directions, we could 
incorporate more covariates, including not only cis-SNPs but also trans-SNPs, in future research. 
We could involve penalty functions when the number of covariates exceeds the number of subjects. 
%Furthermore, we can adopt the false discovery rate (FDR) procedures to control errors for multiple cis-effect tests across genes.

%develop an empirical Bayes approach to estimate precision matrices jointly across multiple tissues. 

%Gaussian graphical models have been applied to infer the relationship between expression levels of genes, where the precision matrix for multivariate normal data has an interpretation of conditional dependence. Most of existing methods for Gaussian graphical models only focus on estimation of a single precision matrix. However, in the GTEx project, gene expression levels are measured over multiple tissues. Estimating precision matrices jointly across multiple tissues has the potential to improve the estimation by borrowing information across tissues. I plan to develop an empirical Bayes approach for joint estimation of multiple relevant precision matrices. Specifically, we can incorporate information across tissues through estimating a common prior on the multiple precision matrices. 

\section*{Acknowledgements}
The authors would like to thank Jianqiao Wang for his help about real data process. This research was supported by NIH grants GM123056 and GM129781, and NSF grant DMS-2210860.

\section*{SUPPLEMENTARY MATERIALS} 
The online Supplemental Materials include EM algorithm for missing data case, additional simulations and proofs of Theorems 1-3. 

%}
\bibliographystyle{imsart-nameyear}
{%\setstretch{1.2} %\footnotesize
	\bibliography{my_bib}

\begin{thebibliography}{19}
% BibTex style file: imsart-nameyear.bst, 2017-11-03
% Default style options (sort=1,type=nameyear).
% Used options (sort=1,type=nameyear).

\bibitem[\protect\citeauthoryear{Bhadra and Mallick}{2013}]{bhadra2013joint}
\begin{barticle}[author]
\bauthor{\bsnm{Bhadra},~\bfnm{Anindya}\binits{A.}} \AND
  \bauthor{\bsnm{Mallick},~\bfnm{Bani~K}\binits{B.~K.}}
(\byear{2013}).
\btitle{Joint high-dimensional Bayesian variable and covariance selection with
  an application to eQTL analysis}.
\bjournal{Biometrics}
\bvolume{69}
\bpages{447--457}.
\end{barticle}
\endbibitem

\bibitem[\protect\citeauthoryear{Brem et~al.}{2005}]{brem2005genetic}
\begin{barticle}[author]
\bauthor{\bsnm{Brem},~\bfnm{Rachel~B}\binits{R.~B.}},
  \bauthor{\bsnm{Storey},~\bfnm{John~D}\binits{J.~D.}},
  \bauthor{\bsnm{Whittle},~\bfnm{Jacqueline}\binits{J.}} \AND
  \bauthor{\bsnm{Kruglyak},~\bfnm{Leonid}\binits{L.}}
(\byear{2005}).
\btitle{Genetic interactions between polymorphisms that affect gene expression
  in yeast}.
\bjournal{Nature}
\bvolume{436}
\bpages{701--703}.
\end{barticle}
\endbibitem

\bibitem[\protect\citeauthoryear{Consortium et~al.}{2017}]{gtex}
\begin{barticle}[author]
\bauthor{\bsnm{Consortium},~\bfnm{GTEx}\binits{G.}} \betal{et~al.}
(\byear{2017}).
\btitle{Genetic effects on gene expression across human tissues}.
\bjournal{Nature}
\bvolume{550}
\bpages{204--213}.
\end{barticle}
\endbibitem

\bibitem[\protect\citeauthoryear{Dempster, Laird and
  Rubin}{1977}]{dempster1977maximum}
\begin{barticle}[author]
\bauthor{\bsnm{Dempster},~\bfnm{Arthur~P}\binits{A.~P.}},
  \bauthor{\bsnm{Laird},~\bfnm{Nan~M}\binits{N.~M.}} \AND
  \bauthor{\bsnm{Rubin},~\bfnm{Donald~B}\binits{D.~B.}}
(\byear{1977}).
\btitle{Maximum likelihood from incomplete data via the EM algorithm}.
\bjournal{Journal of the Royal Statistical Society: Series B (Methodological)}
\bvolume{39}
\bpages{1--22}.
\end{barticle}
\endbibitem

\bibitem[\protect\citeauthoryear{Duong et~al.}{2016}]{duong2016using}
\begin{barticle}[author]
\bauthor{\bsnm{Duong},~\bfnm{Dat}\binits{D.}},
  \bauthor{\bsnm{Zou},~\bfnm{Jennifer}\binits{J.}},
  \bauthor{\bsnm{Hormozdiari},~\bfnm{Farhad}\binits{F.}},
  \bauthor{\bsnm{Sul},~\bfnm{Jae~Hoon}\binits{J.~H.}},
  \bauthor{\bsnm{Ernst},~\bfnm{Jason}\binits{J.}},
  \bauthor{\bsnm{Han},~\bfnm{Buhm}\binits{B.}} \AND
  \bauthor{\bsnm{Eskin},~\bfnm{Eleazar}\binits{E.}}
(\byear{2016}).
\btitle{Using genomic annotations increases statistical power to detect
  eGenes}.
\bjournal{Bioinformatics}
\bvolume{32}
\bpages{i156--i163}.
\end{barticle}
\endbibitem

\bibitem[\protect\citeauthoryear{Duong et~al.}{2017}]{duong2017applying}
\begin{barticle}[author]
\bauthor{\bsnm{Duong},~\bfnm{Dat}\binits{D.}},
  \bauthor{\bsnm{Gai},~\bfnm{Lisa}\binits{L.}},
  \bauthor{\bsnm{Snir},~\bfnm{Sagi}\binits{S.}},
  \bauthor{\bsnm{Kang},~\bfnm{Eun~Yong}\binits{E.~Y.}},
  \bauthor{\bsnm{Han},~\bfnm{Buhm}\binits{B.}},
  \bauthor{\bsnm{Sul},~\bfnm{Jae~Hoon}\binits{J.~H.}} \AND
  \bauthor{\bsnm{Eskin},~\bfnm{Eleazar}\binits{E.}}
(\byear{2017}).
\btitle{Applying meta-analysis to genotype-tissue expression data from multiple
  tissues to identify eQTLs and increase the number of eGenes}.
\bjournal{Bioinformatics}
\bvolume{33}
\bpages{i67--i74}.
\end{barticle}
\endbibitem

\bibitem[\protect\citeauthoryear{Flutre et~al.}{2013}]{Stephens}
\begin{barticle}[author]
\bauthor{\bsnm{Flutre},~\bfnm{Timoth{\'e}e}\binits{T.}},
  \bauthor{\bsnm{Wen},~\bfnm{Xiaoquan}\binits{X.}},
  \bauthor{\bsnm{Pritchard},~\bfnm{Jonathan}\binits{J.}} \AND
  \bauthor{\bsnm{Stephens},~\bfnm{Matthew}\binits{M.}}
(\byear{2013}).
\btitle{A statistical framework for joint eQTL analysis in multiple tissues}.
\bjournal{PLoS Genet}
\bvolume{9}
\bpages{e1003486}.
\end{barticle}
\endbibitem

\bibitem[\protect\citeauthoryear{Gamazon et~al.}{2015}]{gamazon2015gene}
\begin{barticle}[author]
\bauthor{\bsnm{Gamazon},~\bfnm{Eric~R}\binits{E.~R.}},
  \bauthor{\bsnm{Wheeler},~\bfnm{Heather~E}\binits{H.~E.}},
  \bauthor{\bsnm{Shah},~\bfnm{Kaanan~P}\binits{K.~P.}},
  \bauthor{\bsnm{Mozaffari},~\bfnm{Sahar~V}\binits{S.~V.}},
  \bauthor{\bsnm{Aquino-Michaels},~\bfnm{Keston}\binits{K.}},
  \bauthor{\bsnm{Carroll},~\bfnm{Robert~J}\binits{R.~J.}},
  \bauthor{\bsnm{Eyler},~\bfnm{Anne~E}\binits{A.~E.}},
  \bauthor{\bsnm{Denny},~\bfnm{Joshua~C}\binits{J.~C.}},
  \bauthor{\bsnm{Nicolae},~\bfnm{Dan~L}\binits{D.~L.}},
  \bauthor{\bsnm{Cox},~\bfnm{Nancy~J}\binits{N.~J.}} \betal{et~al.}
(\byear{2015}).
\btitle{A gene-based association method for mapping traits using reference
  transcriptome data}.
\bjournal{Nature genetics}
\bvolume{47}
\bpages{1091}.
\end{barticle}
\endbibitem

\bibitem[\protect\citeauthoryear{Gosik et~al.}{2017}]{gosik2017iform}
\begin{barticle}[author]
\bauthor{\bsnm{Gosik},~\bfnm{Kirk}\binits{K.}},
  \bauthor{\bsnm{Kong},~\bfnm{Lan}\binits{L.}},
  \bauthor{\bsnm{Chinchilli},~\bfnm{Vernon~M}\binits{V.~M.}} \AND
  \bauthor{\bsnm{Wu},~\bfnm{Rongling}\binits{R.}}
(\byear{2017}).
\btitle{iFORM/eQTL: an ultrahigh-dimensional platform for inferring the global
  genetic architecture of gene transcripts}.
\bjournal{Briefings in bioinformatics}
\bvolume{18}
\bpages{250--259}.
\end{barticle}
\endbibitem

\bibitem[\protect\citeauthoryear{Gusev et~al.}{2016}]{gusev2016integrative}
\begin{barticle}[author]
\bauthor{\bsnm{Gusev},~\bfnm{Alexander}\binits{A.}},
  \bauthor{\bsnm{Ko},~\bfnm{Arthur}\binits{A.}},
  \bauthor{\bsnm{Shi},~\bfnm{Huwenbo}\binits{H.}},
  \bauthor{\bsnm{Bhatia},~\bfnm{Gaurav}\binits{G.}},
  \bauthor{\bsnm{Chung},~\bfnm{Wonil}\binits{W.}},
  \bauthor{\bsnm{Penninx},~\bfnm{Brenda~WJH}\binits{B.~W.}},
  \bauthor{\bsnm{Jansen},~\bfnm{Rick}\binits{R.}},
  \bauthor{\bsnm{De~Geus},~\bfnm{Eco~JC}\binits{E.~J.}},
  \bauthor{\bsnm{Boomsma},~\bfnm{Dorret~I}\binits{D.~I.}},
  \bauthor{\bsnm{Wright},~\bfnm{Fred~A}\binits{F.~A.}} \betal{et~al.}
(\byear{2016}).
\btitle{Integrative approaches for large-scale transcriptome-wide association
  studies}.
\bjournal{Nature genetics}
\bvolume{48}
\bpages{245--252}.
\end{barticle}
\endbibitem

\bibitem[\protect\citeauthoryear{Hu et~al.}{2019}]{hu2019statistical}
\begin{barticle}[author]
\bauthor{\bsnm{Hu},~\bfnm{Yiming}\binits{Y.}},
  \bauthor{\bsnm{Li},~\bfnm{Mo}\binits{M.}},
  \bauthor{\bsnm{Lu},~\bfnm{Qiongshi}\binits{Q.}},
  \bauthor{\bsnm{Weng},~\bfnm{Haoyi}\binits{H.}},
  \bauthor{\bsnm{Wang},~\bfnm{Jiawei}\binits{J.}},
  \bauthor{\bsnm{Zekavat},~\bfnm{Seyedeh~M}\binits{S.~M.}},
  \bauthor{\bsnm{Yu},~\bfnm{Zhaolong}\binits{Z.}},
  \bauthor{\bsnm{Li},~\bfnm{Boyang}\binits{B.}},
  \bauthor{\bsnm{Gu},~\bfnm{Jianlei}\binits{J.}},
  \bauthor{\bsnm{Muchnik},~\bfnm{Sydney}\binits{S.}} \betal{et~al.}
(\byear{2019}).
\btitle{A statistical framework for cross-tissue transcriptome-wide association
  analysis}.
\bjournal{Nature genetics}
\bvolume{51}
\bpages{568--576}.
\end{barticle}
\endbibitem

\bibitem[\protect\citeauthoryear{Li et~al.}{2018}]{GenLi}
\begin{barticle}[author]
\bauthor{\bsnm{Li},~\bfnm{Gen}\binits{G.}},
  \bauthor{\bsnm{Shabalin},~\bfnm{Andrey~A}\binits{A.~A.}},
  \bauthor{\bsnm{Rusyn},~\bfnm{Ivan}\binits{I.}},
  \bauthor{\bsnm{Wright},~\bfnm{Fred~A}\binits{F.~A.}} \AND
  \bauthor{\bsnm{Nobel},~\bfnm{Andrew~B}\binits{A.~B.}}
(\byear{2018}).
\btitle{An empirical Bayes approach for multiple tissue eQTL analysis}.
\bjournal{Biostatistics}
\bvolume{19}
\bpages{391--406}.
\end{barticle}
\endbibitem

\bibitem[\protect\citeauthoryear{Lonsdale et~al.}{2013}]{lonsdale2013genotype}
\begin{barticle}[author]
\bauthor{\bsnm{Lonsdale},~\bfnm{John}\binits{J.}},
  \bauthor{\bsnm{Thomas},~\bfnm{Jeffrey}\binits{J.}},
  \bauthor{\bsnm{Salvatore},~\bfnm{Mike}\binits{M.}},
  \bauthor{\bsnm{Phillips},~\bfnm{Rebecca}\binits{R.}},
  \bauthor{\bsnm{Lo},~\bfnm{Edmund}\binits{E.}},
  \bauthor{\bsnm{Shad},~\bfnm{Saboor}\binits{S.}},
  \bauthor{\bsnm{Hasz},~\bfnm{Richard}\binits{R.}},
  \bauthor{\bsnm{Walters},~\bfnm{Gary}\binits{G.}},
  \bauthor{\bsnm{Garcia},~\bfnm{Fernando}\binits{F.}},
  \bauthor{\bsnm{Young},~\bfnm{Nancy}\binits{N.}} \betal{et~al.}
(\byear{2013}).
\btitle{The genotype-tissue expression (GTEx) project}.
\bjournal{Nature genetics}
\bvolume{45}
\bpages{580--585}.
\end{barticle}
\endbibitem

\bibitem[\protect\citeauthoryear{Stegle et~al.}{2012}]{stegle2012using}
\begin{barticle}[author]
\bauthor{\bsnm{Stegle},~\bfnm{Oliver}\binits{O.}},
  \bauthor{\bsnm{Parts},~\bfnm{Leopold}\binits{L.}},
  \bauthor{\bsnm{Piipari},~\bfnm{Matias}\binits{M.}},
  \bauthor{\bsnm{Winn},~\bfnm{John}\binits{J.}} \AND
  \bauthor{\bsnm{Durbin},~\bfnm{Richard}\binits{R.}}
(\byear{2012}).
\btitle{Using probabilistic estimation of expression residuals (PEER) to obtain
  increased power and interpretability of gene expression analyses}.
\bjournal{Nature Protocols}
\bvolume{7}
\bpages{500}.
\end{barticle}
\endbibitem

\bibitem[\protect\citeauthoryear{Stouffer et~al.}{1949}]{stouffer1949american}
\begin{bbook}[author]
\bauthor{\bsnm{Stouffer},~\bfnm{Samuel~A}\binits{S.~A.}},
  \bauthor{\bsnm{Suchman},~\bfnm{Edward~A}\binits{E.~A.}},
  \bauthor{\bsnm{DeVinney},~\bfnm{Leland~C}\binits{L.~C.}},
  \bauthor{\bsnm{Star},~\bfnm{Shirley~A}\binits{S.~A.}} \AND
  \bauthor{\bsnm{Williams~Jr},~\bfnm{Robin~M}\binits{R.~M.}}
(\byear{1949}).
\btitle{The american soldier: Adjustment during army life.(studies in social
  psychology in world war ii), vol. 1}.
\bpublisher{Princeton Univ. Press}.
\end{bbook}
\endbibitem

\bibitem[\protect\citeauthoryear{Stranger
  et~al.}{2007}]{stranger2007population}
\begin{barticle}[author]
\bauthor{\bsnm{Stranger},~\bfnm{Barbara~E}\binits{B.~E.}},
  \bauthor{\bsnm{Nica},~\bfnm{Alexandra~C}\binits{A.~C.}},
  \bauthor{\bsnm{Forrest},~\bfnm{Matthew~S}\binits{M.~S.}},
  \bauthor{\bsnm{Dimas},~\bfnm{Antigone}\binits{A.}},
  \bauthor{\bsnm{Bird},~\bfnm{Christine~P}\binits{C.~P.}},
  \bauthor{\bsnm{Beazley},~\bfnm{Claude}\binits{C.}},
  \bauthor{\bsnm{Ingle},~\bfnm{Catherine~E}\binits{C.~E.}},
  \bauthor{\bsnm{Dunning},~\bfnm{Mark}\binits{M.}},
  \bauthor{\bsnm{Flicek},~\bfnm{Paul}\binits{P.}},
  \bauthor{\bsnm{Koller},~\bfnm{Daphne}\binits{D.}} \betal{et~al.}
(\byear{2007}).
\btitle{Population genomics of human gene expression}.
\bjournal{Nature Genetics}
\bvolume{39}
\bpages{1217--1224}.
\end{barticle}
\endbibitem

\bibitem[\protect\citeauthoryear{Sul et~al.}{2013}]{sul2013effectively}
\begin{barticle}[author]
\bauthor{\bsnm{Sul},~\bfnm{Jae~Hoon}\binits{J.~H.}},
  \bauthor{\bsnm{Han},~\bfnm{Buhm}\binits{B.}},
  \bauthor{\bsnm{Ye},~\bfnm{Chun}\binits{C.}},
  \bauthor{\bsnm{Choi},~\bfnm{Ted}\binits{T.}} \AND
  \bauthor{\bsnm{Eskin},~\bfnm{Eleazar}\binits{E.}}
(\byear{2013}).
\btitle{Effectively identifying eQTLs from multiple tissues by combining mixed
  model and meta-analytic approaches}.
\bjournal{PLoS Genet}
\bvolume{9}
\bpages{e1003491}.
\end{barticle}
\endbibitem

\bibitem[\protect\citeauthoryear{Wang et~al.}{2016}]{wang2016imputing}
\begin{barticle}[author]
\bauthor{\bsnm{Wang},~\bfnm{Jiebiao}\binits{J.}},
  \bauthor{\bsnm{Gamazon},~\bfnm{Eric~R}\binits{E.~R.}},
  \bauthor{\bsnm{Pierce},~\bfnm{Brandon~L}\binits{B.~L.}},
  \bauthor{\bsnm{Stranger},~\bfnm{Barbara~E}\binits{B.~E.}},
  \bauthor{\bsnm{Im},~\bfnm{Hae~Kyung}\binits{H.~K.}},
  \bauthor{\bsnm{Gibbons},~\bfnm{Robert~D}\binits{R.~D.}},
  \bauthor{\bsnm{Cox},~\bfnm{Nancy~J}\binits{N.~J.}},
  \bauthor{\bsnm{Nicolae},~\bfnm{Dan~L}\binits{D.~L.}} \AND
  \bauthor{\bsnm{Chen},~\bfnm{Lin~S}\binits{L.~S.}}
(\byear{2016}).
\btitle{Imputing gene expression in uncollected tissues within and beyond
  GTEx}.
\bjournal{The American Journal of Human Genetics}
\bvolume{98}
\bpages{697--708}.
\end{barticle}
\endbibitem

\bibitem[\protect\citeauthoryear{Zeng, Wang and Huang}{2017}]{zeng2017cis}
\begin{barticle}[author]
\bauthor{\bsnm{Zeng},~\bfnm{Ping}\binits{P.}},
  \bauthor{\bsnm{Wang},~\bfnm{Ting}\binits{T.}} \AND
  \bauthor{\bsnm{Huang},~\bfnm{Shuiping}\binits{S.}}
(\byear{2017}).
\btitle{Cis-SNPs set testing and predixcan analysis for gene expression data
  using linear mixed models}.
\bjournal{Scientific reports}
\bvolume{7}
\bpages{1--11}.
\end{barticle}
\endbibitem

\end{thebibliography}
}

\end{document}